\documentclass[a4paper,fleqn]{cas-dc}

\usepackage[numbers]{natbib}

\usepackage{lipsum}
\usepackage{siunitx} 
\usepackage{physics}
\usepackage{amsmath}
\usepackage[version=4]{mhchem} 
\usepackage{graphicx,caption}

\newcommand{\refeq}[1]  {(\ref{#1})}
\newcommand{\reffig}[1]  {Fig.~\ref{#1}}
\newcommand{\Comphy}{{\textit{Comphy}}}

\newcommand{\Er}[1][]{\ensuremath{{E}_{\mathrm{R#1}}}}

\newcommand{\dq}[1][]{\ensuremath{{\Delta Q}}}
\newcommand{\dE}[1][]{\ensuremath{{\Delta E}}}

\newcommand{\sio}[1][]{\ensuremath{\ce{SiO2}}}
\newcommand{\hfo}[1][]{\ensuremath{\ce{HfO2}}}

\def\tsc#1{\csdef{#1}{\textsc{\lowercase{#1}}\xspace}}
\tsc{WGM}
\tsc{QE}
\newcommand{\dVth}[1][]{\ensuremath{{\Delta V_\mathrm{th}}}}

\begin{document}
\let\WriteBookmarks\relax
\def\floatpagepagefraction{1}
\def\textpagefraction{.001}

\shorttitle{Comphy v3.0}    

\shortauthors{D. Waldhoer \textit{et al}}  

\title [mode = title]{\textit{Comphy v3.0} -- A Compact-Physics Framework for Modeling Charge Trapping Related Reliability Phenomena in MOS Devices}


%

\author[1]{Dominic Waldhoer}[orcid=0000-0002-8631-5681]
\cormark[1]{waldhoer@iue.tuwien.ac.at}
\ead{waldhoer@iue.tuwien.ac.at}

\author[1,4]{Christian Schleich}[orcid=0000-0002-8832-520X]
\author[1,4]{Jakob Michl}[orcid=0000-0003-2539-3245]
\author[2]{Alexander Grill}[orcid=0000-0003-1615-1033]
\author[2]{Dieter Claes}[orcid=0000-0002-0356-0973]
\author[1]{Alexander Karl}[orcid=0000-0003-2221-8038]
\author[1]{Theresia Knobloch}[orcid=0000-0001-5156-9510]
\author[3]{Gerhard Rzepa} [orcid=0000-0002-3711-1957]
\author[2]{Jacopo Franco} [orcid=0000-0002-7382-8605]
\author[2]{Ben Kaczer} [orcid=0000-0002-1484-4007]

\author[1,4]{Michael Waltl}[orcid=0000-0001-6042-759X]
\author[1]{Tibor Grasser}[orcid=0000-0001-6536-2238]
\cormark[1]{grasser@iue.tuwien.ac.at}
\ead{grasser@iue.tuwien.ac.at}

\affiliation[1]{organization={Institute for Microelectronics, TU Wien},
            country={Austria}}

\affiliation[2]{organization={imec},	
	city={Leuven},
	country={Belgium}}

\affiliation[3]{organization={Global TCAD Solutions},	
	city={Wien},
	country={Austria}}

\affiliation[4]{organization={Christian Doppler Laboratory for SDS at the Institute for Microelectronics, TU Wien},	
	country={Austria}}

\begin{abstract}
	Charge trapping plays an important role for the reliability of electronic devices and manifests itself in various phenomena like bias temperature instability (BTI), random telegraph noise (RTN), hysteresis or trap-assisted tunneling (TAT). In this work we present \textit{Comphy v3.0}, an open source physical framework for modeling these effects in a unified fashion using nonradiative multiphonon theory on a one-dimensional device geometry. Here we give an overview about the underlying theory, discuss newly introduced features compared to the original \textit{Comphy} framework and also review recent advances in reliability physics enabled by these new features. The usefulness of \textit{Comphy v3.0} for the reliability community is highlighted by several practical examples including automatic extraction of defect distributions, modeling of TAT in high-$\kappa$ capacitors and BTI/RTN modeling at cryogenic temperatures.
\end{abstract}


\begin{highlights}
\item User-friendly Python package providing a compact model for device reliability 
\item Automatic defect parameter extraction and reliability model creation from measure-stress-measure experiments
\item Unified framework for modeling bias temperature instability, random telegraph noise and trap-assisted tunneling
\item Quantum mechanical models to consider nuclear tunneling relevant for charge trapping at cryogenic temperatures
\end{highlights}


\maketitle

\section{Introduction}\label{}
Charge trapping at defects in the gate dielectric or at the oxide/semiconductor interface of a MOSFET device causes various reliability challenges in different applications. For instance, charge trapping due to temperature and/or bias stress induces a drift in the device characteristic, commonly referred to as bias temperature instability (BTI)~\cite{Miura_1966, SCHRODER2007841, STATHIS2018244}, leading to a potential failure in both digital and analog circuits over the course of the device lifetime. In memory applications on the other hand, trap-assisted tunneling (TAT)~\cite{chou1997} through oxide defects can limit the data retention times of the stored information~\cite{8103049}.
In non-standard technologies, on the other hand, the defect densities are typically higher, making charge trapping already relevant during initial operation and can lead to a clearly visible hysteresis in the transfer characteristics~\cite{Rescher2016,Knobloch2022}. 

While the aforementioned phenomena typically exhibit a pronounced temperature activation~\cite{AICHINGER08B,AICHINGER09,ToledanoLuque2011}, which might lead to the conclusion that charge trapping becomes irrelevant at lower temperatures, it has been demonstrated that charge trapping processes also occur at cryogenic temperatures due to nuclear tunneling. One example is the prominence of positive BTI at $\SI{4}{\kelvin}$ observed in high-$\kappa$ gate stacks~\cite{Michl2021EfficientModelingOf2}. Another manifestation of nuclear tunneling is the occurrence of random telegraph noise (RTN)~\cite{PhysRevLett.52.228} in control circuits for solid-state qubits, which causes a loss of quantum coherence~\cite{PhysRevA.89.012330, PhysRevB.105.245413}. 

Besides these detrimental effects, charge trapping can also be leveraged for certain applications, e.g. to store information in charge trap flash devices~\cite{choi2020} or as physical unclonable function by providing a unique device-specific noise signature~\cite{chen2015}.

These examples emphasize the importance of understanding the microscopic nature of defects as well as the need to accurately describe charge trapping in devices. Modern TCAD simulation packages like Minimos-NT~\cite{MINIMOS-NT,minimos} or Sentaurus Device~\cite{sentaurus} include a multi-state nonradiative multiphonon (NMP) model~\cite{GRASSER201239} to incorporate the effects of charge trapping on the device performance and lifetime. However, such models are often too complex to be well calibrated with the available experimental data. Furthermore, an overly elaborate modeling approach can distract from the essential physics of charge trapping, which, in many relevant cases, can be well described within a simplified 2-state NMP model~\cite{rzepa2017}. In this spirit, we recently introduced the \textit{Comphy} (compact-physics) framework~\cite{rzepa2018comphy}, a light-weight Python package, which is designed to simulate various charge-trapping related reliability issues based on the 2-state NMP model within a compact 1D device simulator.

While \textit{Comphy} in its original form was intended as a proof-of-concept to demonstrate the feasibility of accurate and efficient reliability modeling of core logic FETs, it is now frequently employed to assess and understand the degradation in experimental devices, ranging from novel memory applications~\cite{sullivan2020} to power electronics~\cite{schleich_iedm} in silicon but also other semiconductors such as SiC.
Although accurate reliability models are valuable in their own right for device lifetime predictions, the physics-based modeling approach used in \textit{Comphy} also allows to draw conclusions about the atomistic nature of defects responsible for charge trapping by comparison of model parameters to theoretical predictions based on density functional theory (DFT)~\cite{goes2018,Waldhoer2020}. Furthermore, a physical model can provide clues for possible design improvements. For example, based on \textit{Comphy} predictions, a dipole layer introduced at the SiO$_2$/HfO$_2$ interface of high-$\kappa $ gate stacks was demonstrated to improve BTI by suppressing the defect-carrier interaction~\cite{Franco2013,franco2018}.

In this paper we summarize our recent efforts to further extend the original \textit{Comphy} framework towards a powerful and user-friendly reliability code, culminating in the publically available \textit{Comphy v3.0} package~\cite{comphy3}.








\section{Models and Features}
Compared to \textit{Comphy v1.0}, several newly developed models and features have been added in order to meet the current demands for nanoscale device modeling. In the following we give a brief overview of the most relevant new features in \textit{Comphy v3.0}, a detailed description is then provided in later sections of this paper.

\subsection{Parameter Extraction Methods}
One of the main motivations behind a framework like \textit{Comphy} is to obtain a physically-motivated degradation model based on experimental data points at accelerated stress conditions. A well-calibrated model can then be used to estimate the device lifetime by extrapolating the degradation to typical use conditions. As most gate oxides are amorphous thin films, they show a wide distribution of defect parameters due to the varying local chemical environment of the defects. Furthermore, most degradation experiments are conducted on large-area devices where only the collective response of a whole defect ensemble can be observed. Hence, the underlying defect parameter distributions have to be inferred from these observations. Previous studies mostly assumed Gaussian distributions for the parameters~\cite{rzepa2018comphy,POBEGEN13E} and their mean and sigma values were obtained from a non-linear optimization procedure by fitting experimental data. However, such a procedure can lead to extraction artifacts because it enforces a certain shape of the distribution. Furthermore, the optimization requires a good initial guess, becomes more tedious and requires frequent manual intervention particularly when multiple defect bands are involved, e.g. interacting electron and hole traps as observed in SiC/SiO$_2$ devices~\cite{schleich_iedm,afanasev1997}. In order to circumvent these issues, \textit{Comphy v3.0} offers a novel method of parameter extraction named \textit{Effective Single Defect Decomposition} (ESiD)~\cite{waldhoer2021}, which allows for a semi-automatic extraction of defect parameters from experimental measure-stress-measure traces without the aforementioned assumptions about their distribution and, in particular, without requiring a good initial guess.

\subsection{Trap-Assisted Tunneling (TAT)} While the most pronounced effects of charge trapping at defects in the oxide  are electrostatic shifts manifesting as BTI, RTN and hysteresis, the same mechanism can also facilitate a parasitic gate leakage current by conductance (``hopping'') over defects~\cite{chou1997}. Contrary to direct (DT) or Fowler-Nordheim (FN) tunneling~\cite{tsu1973tunneling}, the resulting currents exhibit a strong temperature dependence, indicating a charge transfer mediated by a multiphonon process similar to charge trapping.  Although several different models have been proposed in  literature to describe this phenomenon \cite{schenk1995, larcher2003, zhang2011}, it is usually treated separately from charge trapping in the context of BTI. \textit{Comphy v3.0} includes our recently developed unified approach~\cite{schleich2022singleI} for TAT and BTI, where both are treated on the same footing within the NMP framework. This approach allows to obtain a consistent defect parameter set for the defect-channel interaction which is then transformed to a corresponding parameter set describing the defect-defect interaction.

\subsection{Charge Trapping at Cryogenic Temperatures} For applications at room temperature and above, the full quantum mechanical NMP model can be reasonably well approximated by its classical analog, the celebrated Marcus theory for charge transfer reactions~\cite{marcus1993}. This model is preferable from a computational point of view, since it only requires calculating the classical transition barrier instead of all vibrational overlaps in NMP theory. However, the classical model predicts a complete freeze-out of charge transfer towards cryogenic temperatures, whereas it is well known that charge trapping and its resulting effects on the device, i.e. BTI and RTN, can be observed at lower temperatures~\cite{Michl2021EfficientModelingOf2,scofield2000,Knobloch2020AnalysisOfSingle, Michl2021EvidenceOfTunneling}. In this regime, the defect reconfiguration upon charge transfer is no longer temperature activated but rather dominated by \textit{nuclear tunneling}. In order to efficiently model charge trapping under these conditions, e.g. for studying RTN in emerging quantum information applications, a Wentzel–Kramers–Brillouin (WKB) approximation to the full quantum mechanical model has been developed~\cite{Michl2021EfficientModelingOf1} and incorporated into \textit{Comphy v3.0}. This approach remedies the computational overhead associated with the quantum mechanical description while remaining sufficiently accurate, hence allowing to simulate a whole ensemble of defects in the nuclear tunneling regime.



\subsection{Coupling to TCAD}
While \textit{Comphy} provides a fast way to assess device reliability, an inherent limitation of the employed 1D geometry is related to the charge sheet approximation (CSA) which is used to model the impact of oxide charges on device electrostatics~\cite{ROUXDITBUISSON1992}.
Even for ideal planar devices, this approximation fails to describe the distribution of threshold voltage shifts caused by the charges of individual traps, furthermore the deviations worsen in the subthreshold regime, where current percolation paths are formed, for example through random discrete dopants (RDD) and random trapped charges~\cite{KACZER16}.
In the case of more complex device geometries such as FinFETs or gate-all-around (GAA) FETs, the modeling of transmission coefficients and the strongly inhomogeneous electric fields and carrier concentrations lead to further inaccuracies in the 1D approximation.

An efficient simulation method to go beyond this 1D approach and to accurately describe the variability-aware reliability of 3D structures is to couple TCAD simulations with \textit{Comphy}. In this approach, the 3D electrostatics are simulated with TCAD for all bias and temperature conditions which are relevant for a given stress scenario to obtain the quantities required by \textit{Comphy} to compute the defect occupancies: The local lattice temperature, the electrostatic potential, the transmission coefficients to all charge reservoirs, as well as the carrier concentrations, energies, and effective masses of the semiconductor and the metal gate.
In addition, the perturbative impedance field method (IFM)~\cite{ELSAYED12} can be applied on top of the quasi-stationary TCAD solutions to accurately model the impact of the defect charges on the device characteristics instead of using the CSA.

With this method, the efficient \textit{Comphy} simulations can leverage full 3D TCAD accuracy as recently demonstrated for a reliability and variability aware design technology co-optimization (DTCO) study of FinFET and nanosheet devices, including BTI degradation of ring-oscillators for 10 years of AC operation~\cite{RZEPA21}.

\section{Theory}
Due to the strong impact on electronic devices, charge trapping at oxide defects is the subject of numerous experimental and theoretical investigations to this day. While RTN is evidently caused by charge trapping~\cite{PhysRevLett.52.228, fleetwood2002}, a connection between oxide defects and BTI has long been suspected \cite{BREED75, HUARD06} but has only recently been firmly established~\cite{grasser_paradigm,GRASSER15B}. Although early theoretical modeling attempts for BTI assumed a reaction-diffusion or reaction-dispersive transport of hydrogen as underlying cause~\cite{jeppson1977,houssa2005,krishnan2004}, elaborate recent measurements show that such a mechanism is fundamentally incompatible with the experimental evidence~\cite{Schanovsky2014,GRASSER14D}. In addition, the parameters used in the reaction-diffusion model, e.g. those related to the properties of hydrogen, are in glaring contradiction to both independent experiments and theoretical expectations~\cite{STATHIS2018244}. Instead, a reaction-limited process involving the phonon-mediated charge transfer to oxide defects is now widely accepted as the dominant cause of BTI~\cite{grasser_paradigm}. This conclusion is mainly based on the pronounced bias- and temperature dependence observed in nanoscale devices \cite{GRASSER14D} as well as the observation that BTI and RTN are  two manifestations of the same microscopic mechanism at shorter time scales~\cite{grasser2014}. At longer time scales additional interface defects are created triggered by a gate-sided hydrogen release mechanism~\cite{GRASSER15B}. Furthermore, recent atomistic studies based on density functional theory support the involvement of defects in BTI for \sio~\cite{grasser_iedm} and \hfo~\cite{alam2021}, the most widely used gate dielectrics. 

Since charge trapping is an essential part of reliability physics, the fundamentals of the underlying  nonradiative multiphonon (NMP) theory, its limitations for reliability modeling as well as  approximations necessary for its implementation in \textit{Comphy} are discussed in the following.

\subsection{Defect States}
In order to model the dynamics of a defect in a device, details about the atomic configurations and the pathways between them are often too complicated to be dealt with explicitly.
Following a certain transition, a defect typically relaxes into its new equilibrium configuration in a matter of picoseconds~\cite{schanovsky_thesis}. This implies that the properties of a defect and its possible future pathways only depend on the current defect state. In other words, the defect is treated as a memory-less system. Using this assumption, the defect states and the transition rates between different states form a Markov chain. All the defect physics, i.e. temperature and bias dependence, is  encoded in the transition rates $k_{ij}$  from an initial state ${i}$ to a final state ${j}$. The probabilities $P_j$ of finding a defect in a certain state at a given time evolve according to the Master equations
\begin{equation}\label{eq:master}
	\frac{\mathrm{d} P_j}{\mathrm{d}t}=\sum_{i\neq j}k_{ij} P_i(t)-k_{ji}P_j(t)\quad \text{with}\quad \sum_{i}P_i(t)=1\,.
\end{equation}
This set of coupled linear differential equations can be solved analytically for an arbitrary number of interacting states using the matrix exponential approach. 

\begin{figure}[t]
	\centering
	\includegraphics[width=\columnwidth]{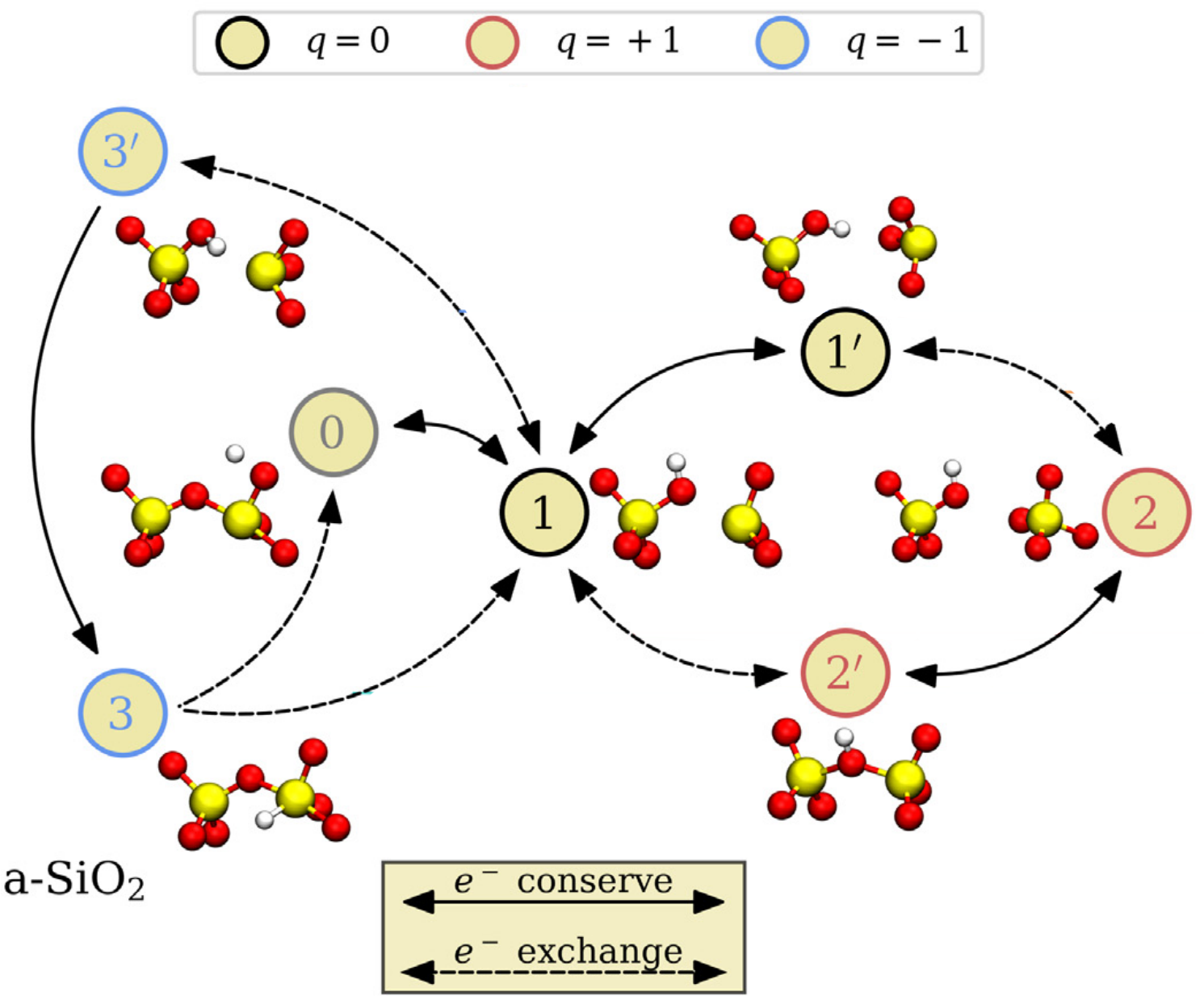}
	\rule{\columnwidth}{1pt}\vspace*{0.3cm}
	\includegraphics[width=\columnwidth]{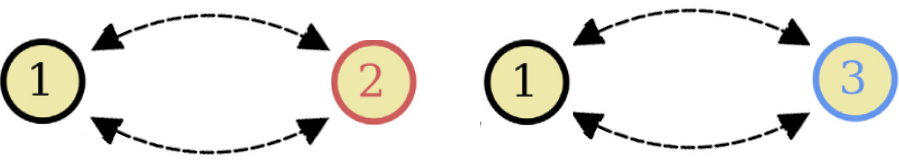}
	\caption{\textbf{Top:} State diagram of the amphoteric hydroxyl E$'$ defect in SiO$_2$. Notice how a single amphotheric defect can capture either electrons or holes by going through its various states. Reprinted from~\cite{WILHELMER2022114801}.  \textbf{Bottom:} Simplified 2-state models for hole and electron trapping used in \textit{Comphy}. Note that in this approximation electron and hole traps become independent.}
	\label{fig:hb_states}
\end{figure}

Time-dependent defect spectroscopy (TDDS)~\cite{PhysRevB.82.245318} and RTN experiments have unambiguously shown that defects can not only undergo charge transitions, but also relaxations to meta-stable configurations within the same charge state. Prime examples for the existence of such meta-stable states are observations of anomalous  RTN~\cite{kirton1989} and so-called switching traps~\cite{lelis}, which show an unusual bias-dependence in their emission time constants, as explained in Sec.~\ref{sec:switching}. The existence of these states was also  established using paramagnetic resonance studies~\cite{lelis,POINDEXTER95} and formed the basis for the development of the 4-state NMP model~\cite{GRASSER201239} currently implemented in commercial TCAD software like Minimos-NT or Sentaurus Device. While this model can give a comprehensive description of hole trapping in Si/\sio\,devices during negative BTI (NBTI) stress, treatment of the less pronounced electron trapping during positive BTI (PBTI) was still lacking. As recently has been demonstrated~\cite{WILHELMER2022114801}, defects in amorphous \sio\, are amphoteric, i.e. they can act as both hole and electron traps. The resulting 7-state diagram is depicted in \reffig{fig:hb_states} for the case of the hydroxyl E$'$ (HE) defect. Although the defect dynamics in such a model are quite complex, its behavior would still be described by \refeq{eq:master}. However, for a proper parameterization this model would require extensive experimental data based on single-defect characterizations, which is exceedingly hard to obtain.

For this reason, \textit{Comphy} condenses the defect dynamics into effective 2-state models for electron and hole traps separately, as shown in \reffig{fig:hb_states}(bottom). Such a simplified model accurately captures the essential physics of charge trapping in most scenarios~\cite{rzepa2017}. Furthermore, this model can be easily parameterized with only a few extended measure-stress-measure (eMSM)~\cite{kaczer2008} sequences using the novel ESiD approach implemented in \textit{Comphy v3.0}, as will be discussed in Sec.~\ref{sec:esid}. Besides these advantages, \refeq{eq:master} has a simple analytical solution
\begin{equation}\label{eq:2state}
P_i(t)=P_i(\infty)+(P_i(0)-P_i(\infty))\cdot \exp\left(-\frac{t}{\tau}\right)
\end{equation}
with the time constant $\tau=1/(k_{ij}+k_{ji})$ 
and the equilibrium solution $P_i(\infty)=\tau k_{ji}.$
This is exploited to efficiently evaluate the occupancies of defects analytically for arbitrary $V_\mathrm{G}$ and $T$ profiles. Note, however, that this approach is only valid for non-interacting defects, an assumption that must be dropped in fully self-consistent calculations or to describe phenomena like multi-trap TAT, see Sec.~\ref{sec:tat} for more details.

\subsection{Limitations of the 2-State Defect Model}
While the effective 2-state model is well suited to reproduce the averaged response of a large ensemble of defects, it is worth discussing its limitations. The following subsections summarize the features introduced by these additional states that cannot be captured by the 2-state model. Whether any of these effects are relevant for a certain application can be checked by comparison to the full 4-state model available in TCAD solutions.

\subsubsection{Switching Traps}\label{sec:switching}
First, the transition between states 2 and 1$'$ as depicted in \reffig{fig:hb_states} leads to the change of the charge state observed in switching traps. 
Essentially, upon hole capture, the defect goes from the neutral state 1 into the positively charged state 2. This transition occurs via the metastable state 2$'$ which becomes clear when looking at atomic configurations of the defect states obtained from DFT. However, after becoming positively charged, many defects have a secondary trap level, typically inside the Si bandgap, which allows them to become neutralized while still in the secondary configuration 1$'$. Note that while a transition to 1$'$ electrically neutralizes the defect, it does not release the distortion at the defect site as 1$'$ is separated by a large barrier from the relaxed neutral configuration 1~\cite{goes2018}. Contrary to the time constants linking 1 and 2 over 2$'$, which are typically large, the time constants linking 2 and 1$'$ are typically short, meaning that these transitions are fast. This allows for a clear experimental distinction: without a transition to 1$'$, the defect would appear like a fixed positive charge which does not react to changes of the Fermi-level, at least up to the point where a transition back to state 1 is favored. On the other hand, with a transition to 1$'$ it appears like the transition from 1 to 2 has created an electrically active defect, a switching trap, which can quickly react to changes in the Fermi-level. As an example, this means that these defects can lead to a  subthreshold slope change and show up as defects in charge pumping (CP) and capacitance-voltage (CV) measurements.

Another interesting consequence is the following: during BTI experiments the gate voltage is typically switched from a stress to a read-out or recovery voltage. If this recovery voltage results in a Fermi-level below the secondary trap level, all these defects will be positively charged and contribute to 100\% to the threshold voltage shift. However, if the Fermi-level during read-out is positioned at the mean of the distribution of the secondary trap level, only 50\% of these traps will be positively charged and thus visible in the threshold voltage shift. This strong read-out bias-dependence can be easily observed experimentally and originally triggered the extension of the 2-state model to a 3-state model (states 1,2, and 1$'$), see \cite{GRASSER09}.

\begin{figure}[t]
	\centering
	\includegraphics[width=\columnwidth]{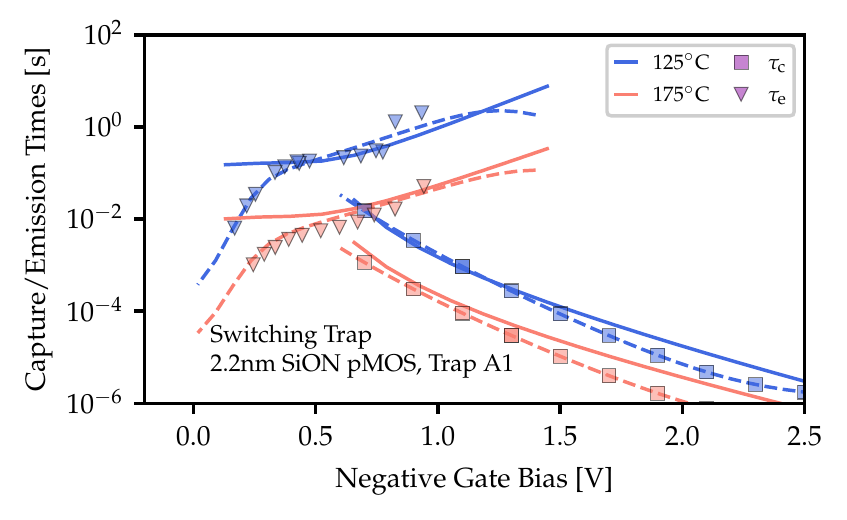}
	\caption{2-state \textbf{(solid)} and 4-state \textbf{(dashed)} model fits to the capture and emission times of a switching trap obtained from TDDS in~\cite{grasser_iedm}. The rapidly decreasing emission times of this switching trap below the threshold voltage ($V_\mathrm{G}\approx \SI{-0.5}{\volt}$) cannot be explained within the 2-state model.}
	\label{fig:2state_4state}
\end{figure}

Additionally, if the transition 2 to 1$'$ leads to a relaxation of the defect back to state 1 via state 1$'$, the emission time will be dramatically reduced, as shown in \reffig{fig:2state_4state}. One consequence of this transition is that such ``positive charge'' can be easily neutralized and the defects restored back to state 1 by switching the transistor into accumulation.  Such an accumulation pulse can be used to experimentally extract the emission time constant under these bias conditions \cite{GRASSER13}.

\subsubsection{Metastable State 2$'$}
Although it has long been known that transitions between states 1 to 2 occur via a metastable state 2$'$~\cite{POINDEXTER95}, the necessity of including this transition into the model was originally deduced from the  stronger temperature activation of the emission time  compared to the capture time constant~\cite{GRASSER10}. Later studies demonstrated that state 2$'$ also leads to a frequency dependence of the capture time constant since at higher frequencies back-transitions from 2$'$ to 1 slow down the overall charge capture ~\cite{GRASSER12C}.

\subsubsection{Precursor Activation}
Another interesting property of oxide defects observed in TDDS experiments is their volatility: they can disappear without a trace just to randomly reappear weeks or months later with exactly the same properties ~\cite{GRASSER13B}. This volatility was interpreted as a transition into the precursor state 0, e.g. by releasing hydrogen from the defect site. The hydrogen is then free to move around, potentially activating new defects nearby via a transition from 0 to 1, or even depassivating Si-H bonds at the interface. Note that contrary to the assumptions of the reaction-diffusion model, the diffusivity of H is very high~\cite{STATHIS2018244}. Thus, H moves around very quickly, not limiting any subsequent reactions. An interesting consequence of the H released in the insulators is that it is available to trigger further reactions. In fact, it has long been known that the Si-H bond is too strong to easily release its H and that the reaction Si-H + H $\rightarrow$ Si-$^*$ + H$_2$ is the only feasible reaction~\cite{STATHIS2018244,cartier1993, cartier1994, HOUSSA07}. Note that the depassivation could also proceed via a proton instead of neutral hydrogen~\cite{rashkeev2001}. Assuming that hydrogen is trapped at the gate side in H-E$'$ centers, they would mostly be positively charged (state 2). Upon application of a bias, they could be neutralized (transition to state 1) or release their H (transition to state 0). This idea led to the gate-sided hydrogen release model~\cite{GRASSER15B} and results in a coupling between charge trapping in the oxide and interface trap creation~\cite{GRASSER09B}. While providing a physical explanation, this model in its current stage is computationally demanding and suffers from numerical instabilities, hence in \textit{Comphy} the creation of interface states is modeled via an empirical double-well barrier~\cite{rzepa2018comphy}.

\subsection{Nonradiative Multiphonon Transitions}

\begin{figure}[t]
	\centering
	\includegraphics[width=\columnwidth]{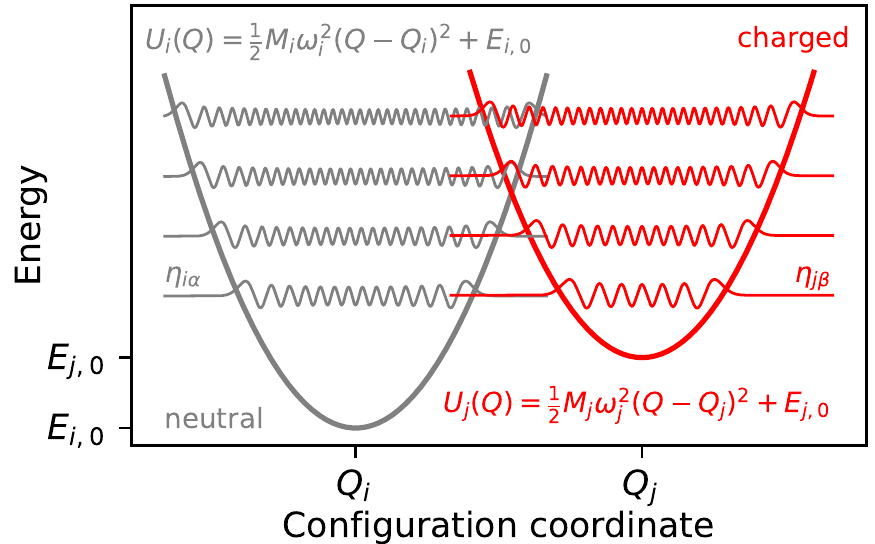}
	\caption{Potential energy curves of a neutral and charged defect state within the harmonic approximation. The overlap integrals of the vibrational wavefunctions $\eta_{i\alpha}$ and $\eta_{j\beta}$ determine the nonradiative transition rates.}
	\label{fig:2state_NMP}
\end{figure}

While mathematically the defects are modeled as Markov chains, the transition rates $k_{ij}$ containing the physical details within this model still have to be determined. As indicated in \reffig{fig:hb_states}, there are two fundamentally different types of transitions. First, there are so-called thermal transitions (solid lines) between states with the same charge. In such cases, the transition rates are determined by a reaction barrier and can be evaluated within the classical transition state theory. Since these transitions are not treated explicitly in the simplified Comphy model, we will not discuss them further here and refer the interested reader to the literature~\cite{nitzan2006chemical}. The second type of transitions involves a charge transfer event typically accompanied by the emission or absorption of multiple phonons from the environment. These transitions are governed by the nonradiative multiphonon formalism~\cite{huang1950,henry1977} which will be outlined in the following. A detailed description can be found in~\cite{goes2018}.

\begin{figure}[t]
	\centering
	\includegraphics[width=\columnwidth]{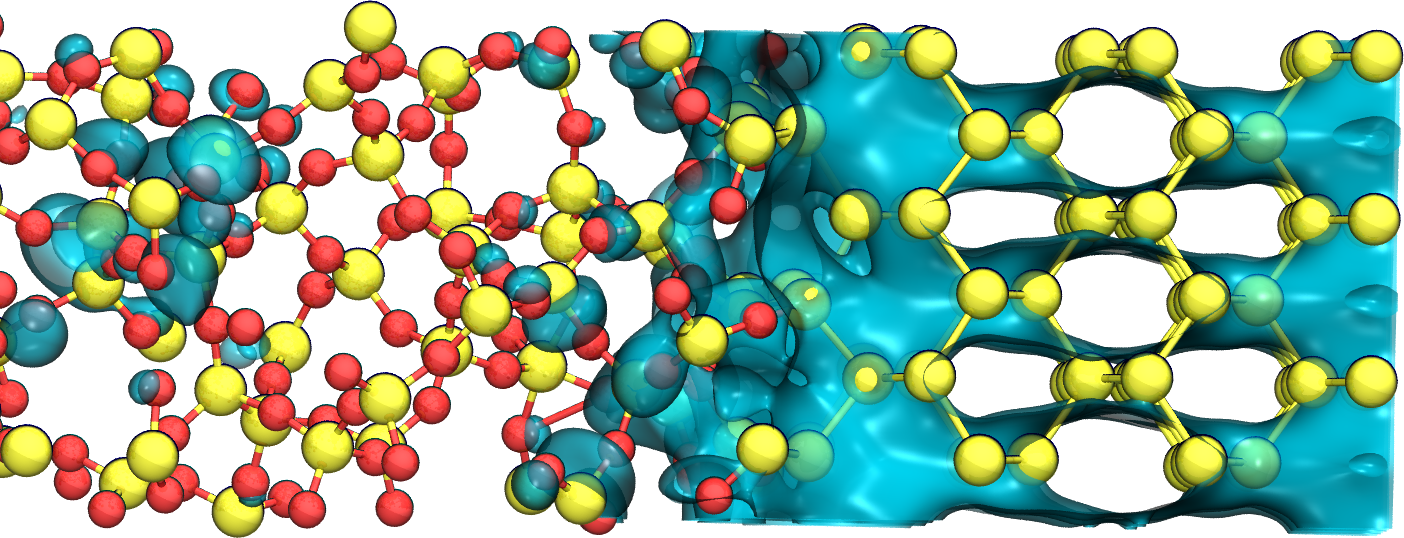}
	\caption{DFT calculations of an atomistic Si/SiO$_2$ interface. \textbf{Left:} Localized defect state in the SiO$_2$. \textbf{Right:} Delocalized wavefunction in the silicon substrate penetrating into the oxide.}
	\label{fig:interface}
\end{figure}

During a charge transition, typically the atomic configuration of the defect changes due to structural relaxations. After decoupling the electronic and vibrational degrees of freedom by applying the Born-Oppenheimer approximation, the system is described by potential energy curves (PECs) $U_i(Q)$ and $U_j(Q)$ for both involved charge states along a so-called reaction coordinate $Q$ as depicted in \reffig{fig:2state_NMP}. Usually the PECs are assumed to be parabolic, i.e. the defect states are approximated as harmonic oscillators. Each PEC belongs to a different diabatic electronic state, e.g. the carrier is either delocalized in the semiconductor substrate or localized at the oxide defect. The corresponding electronic wavefunctions $\Psi_i$ are shown in \reffig{fig:interface}.  On the other hand, the phonons involved in the charge transition are described by the vibrational wavefunctions $\eta_{i\alpha}$ and $\eta_{j\beta}$ arising from the PECs.

In order to obtain the total rate for a transition $i\to j$, all possible transitions between initial and final vibrational states have to be considered. Assuming a weak coupling strength between the defect and the device substrate, the individual vibrational transition rates are given by Fermi's Golden Rule within time-dependent perturbation theory. By assuming a thermal equilibrium in the initial state and by invoking the Franck-Condon principle~\cite{PhysRev.28.1182}, the total transition rate can  be expressed as~\cite{goes2018}
\begin{equation}\label{eq:nmp_rate}
	k_{ij}=A_{ij} f_{ij}
\end{equation}
with the electronic matrix element
\begin{equation}\label{eq:el_matrix}
A_{ij}=\frac{2\pi}{\hbar}|\langle \Psi_i|\hat{H}_\mathrm{el}|\Psi_j\rangle|^2
\end{equation}
and the Franck-Condon lineshape function
\begin{equation}\label{eq:fc_lsf}
f_{ij}=\frac{1}{Z} \sum_{\alpha,\beta} |\langle \eta_{i\alpha}|\eta_{j\beta} \rangle |^2 \exp\Big(-\frac{E_{i\alpha}}{k_\mathrm{B}T}\Big) \delta(E_{i\alpha}-E_{j\beta})\,.
\end{equation}
Here, $\hat{H}_\mathrm{el}$ is the Hamiltonian of the electronic subsystem,
$Z$ is the canonical partition function of the initial state, $E_{i\alpha}$ and $E_{j\beta}$ are the vibrational eigenenergies of the initial and final states respectively, as shown in \reffig{fig:2state_NMP}. While \refeq{eq:nmp_rate} describes the full quantum mechanical NMP transition rates, the usage in device simulations requires further approximations. 

First, the electronic matrix element $A_{ij}$ cannot be calculated within a classical device simulator, since the electronic wavefunctions are not available. Even when employing actual atomistic simulations~\cite{PhysRevApplied.11.044058} based on DFT, accurately obtaining $A_{ij}$ remains highly challenging. However, due to the localized nature of the defect wavefunction and the exponentially decaying substrate wavefunction into the oxide, $A_{ij}$ can be reasonably approximated by a classical capture cross section $\sigma_0$ from kinetic gas theory modified by a WKB based tunneling factor, yielding
\begin{equation}
A_{ij}\propto v_\mathrm{th} \sigma_0 \vartheta_\mathrm{WKB},
\end{equation}
with $v_\mathrm{th}=\sqrt{3 k_\mathrm{B}T/\pi m^*}$ being the thermal velocity of carriers in the substrate. The tunneling factor $\vartheta_\mathrm{WKB}$ is evaluated similarly to \refeq{equ:WKB_full}, as discussed in Sec.~\ref{sec:tsu_esaki}. All results shown in this work assumed a capture cross section of $\SI{e-15}{\square\centi\meter}$.

\begin{figure}[t]
	\centering
	\includegraphics[width=\columnwidth]{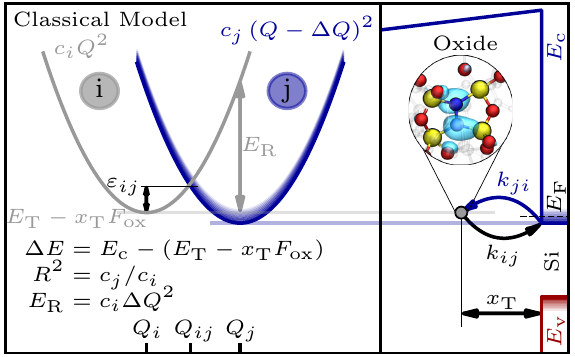}		
	\caption{Classical (high temperature) limit of the NMP model. Here, the transition rate is only determined by the classical barrier $\varepsilon_{ij}$. The PECs and their relative position are defined by the trap level $E_\mathrm{T}$, the relaxation energy $E_\mathrm{R}$ and the curvature ratio $R$. Additionally, the position of the trap ($x_\mathrm{T}$) together with the applied electric field $F_\mathrm{ox}$ across the oxide  introduces a bias-dependent energy offset $\Delta E$.}
	\label{fig:classical_model}
\end{figure}

Second, evaluating all vibrational overlap integrals in \refeq{eq:fc_lsf} is computationally very expensive and hence not suitable for an efficient model. Instead, we employ the classical limit ($T\to \infty$), in which the lineshape function is dominated by the classical barrier $\varepsilon_{ij}$, given by the crossing point of the two PECs, see also \reffig{fig:classical_model}. In this case, the lineshape function can be approximated by
\begin{equation}
f_{ij}\approx \xi \exp\Big(-\frac{\varepsilon_{ij}}{k_\mathrm{B}T}\Big),
\end{equation}
with $\xi$ being a prefactor depending on the exact shape of the PECs. However, this prefactor can usually be ignored due to the dominance of the Arrhenius-like temperature activation. While this approximation gives good results at room temperature and above~\cite{goes2018}, it is not suitable for cryogenic applications. In such cases, a different approach based on a WKB approximation of the vibrational wavefunctions is used, see Sec.~\ref{sec:cryo} for details.

Although not explicitly stated, the quantities $A_{ij}$ and $f_{ij}$ both depend on the energy $E$ of the carrier being exchanged between defect and substrate. Hence \refeq{eq:nmp_rate} has to be integrated over the whole conduction and valence band of the substrate semiconductor in order to obtain the full transition rates. However, this integration can only be done numerically and is hence associated with large computational costs. To circumvent this, we employ the so-called band-edge approximation~\cite{rzepa2018comphy}, which assumes that all the available carriers are located exactly at the conduction or valence band edge. Under this assumption the classical transition rates can be expressed analytically and, in the case of an electron trap interacting with the semiconductor conduction band edge, are given by
\begin{align}
\label{eq:NMPrates_2state_cl}
\begin{split}
k_{ij}^\mathrm{cl}&= n v_\mathrm{th,n}\sigma_{0,n}\vartheta_\mathrm{n}\mathrm{e}^{-\beta({\varepsilon_{ij}-E_\mathrm{F}+E_\mathrm{CB}})}\\
k_{ji}^\mathrm{cl}&= n v_\mathrm{th,n}\sigma_{0,n}\vartheta_\mathrm{n}\mathrm{e}^{-\beta{\varepsilon_{ji}}}
\end{split}
\end{align}
Note however, that for some special cases like the weak electron-phonon coupling regime, the correct value of the barrier $\varepsilon_{ij}$ can deviate from the definition given in \reffig{fig:classical_model}. Details about this intricacy can be found in the original \textit{Comphy} paper~\cite{rzepa2018comphy}, in particular Tab.~4.

\subsection{Defect Parameters}


In order to simulate the threshold voltage shift $\Delta V_\mathrm{th}$ caused by a (partially) charged defect ensemble, the PECs of the individual defects have to be parameterized. As depicted in \reffig{fig:classical_model}, within the harmonic approximation the PECs of the two defect states can be described by
\begin{equation}
U_i(Q)=c_i Q^2\quad \text{and}\quad U_j(Q)=c_j (Q-\Delta Q)^2+\Delta E\,.
\end{equation}
While the PECs can be directly parameterized by the curvatures $c_{i/j}$ at their respective minima, it is more common to use the \textit{relaxation energy}
\begin{equation}
E_\mathrm{R}=U_i(\Delta Q)-U_i(0)=c_i \Delta Q^2
\end{equation} 
and the \textit{curvature ratio} $R=\sqrt{{c_j}/{c_i}}$
instead. Note that the relaxation energy was denoted as $S$ in previous works, however this was changed to $E_\mathrm{R}$ to avoid possible confusion with the Huang-Rhys factor. The energetic offset $\Delta E$ between the two parabolas is determined by the intrinsic trap level of the defect $E_\mathrm{T}$ as well as the applied electric field $F_\mathrm{ox}$ in the oxide and is given by
\begin{equation}
\Delta E=E_\mathrm{C}-E_\mathrm{T}-x_\mathrm{T} F_\mathrm{ox}
\end{equation}
for a charge-free dielectric. Note that such a non-self-consistent treatment is only valid if the charges accumulating in the oxide defects cause only a small shift in the device electrostatics. Otherwise a self-consistent approach solving the Poisson equation including the oxide charges is necessary.

For the defect PECs and their relative position with respect to each other to be uniquely defined, one has to specify the parameter tuple 
\begin{equation}\label{eq:tupel}
\boldsymbol{p}=\left(E_\mathrm{T}, E_\mathrm{R}, R, x_\mathrm{T}, \Delta Q\right)\,.
\end{equation}
Note that in the classical limit the transition rates are only determined by the energetic barrier defined by the PEC crossing point. Hence, the displacement $\Delta Q$ along the configuration coordinate is immaterial in this case. However, in a quantum mechanical treatment, $\Delta Q$ is an important defect parameter determining the likelihood of nuclear tunneling.

\subsubsection*{Parameter Correlations}

\begin{figure}[t]
	\centering
	\includegraphics[width=\columnwidth]{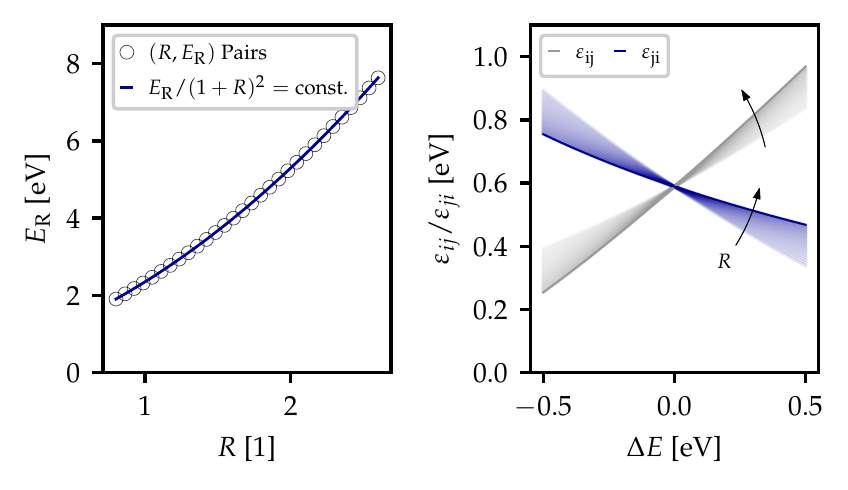}
	\caption{Cross-correlation between the parameters $R$ and $E_\mathrm{R}$. When modified along the correlation curve \textbf{(left)}, $R$ and $E_\mathrm{R}$ can vary widely, however the resulting classical barriers remain rather similar \textbf{(right)}.}
	\label{fig:corel}
\end{figure}

While such a tuple $\boldsymbol{p}$ is sufficient to determine the transition rates of a defect and hence also its capture ($\tau_\mathrm{c}$) and emission times ($\tau_\mathrm{e}$), for the inverse problem, i.e. extracting defect parameters from a set of experimental $\Delta V_\mathrm{th}$ curves, one has to check for cross-correlations between the model parameters. Framed differently, the question is, whether or not a \textit{unique} set of defect parameters can be extracted from the available experimental data obtained with electrical characterization methods. 

Considering only the classical limit of the NMP model, we have previously identified a non-linear cross-correlation between the parameters $R$ and $E_\mathrm{R}$~\cite{waldhoer2021}. The classical barrier can be expanded around $\Delta E=0$, yielding
\begin{equation}
\varepsilon_{ij}(\Delta E)=\frac{E_\mathrm{R}}{(1+R)^2}+\frac{R}{1+R}\Delta E+\mathcal{O}(\Delta E^2)\,.
\end{equation} 
As illustrated in \reffig{fig:corel}, we observe that keeping the zeroth order term in this expansion fixed, $R$ and $E_\mathrm{R}$ can vary over a wide parameter range resulting in very similar barriers. However, since the barrier is the defining feature for the electrically measurable $\Delta V_\mathrm{th}$ response in the classical model, this implies that simultaneously extracting both $R$ and $E_\mathrm{R}$ uniquely from experiments is challenging.

We suspect this $(R,E_\mathrm{R})$ correlation to be responsible for the unphysically high relaxation energies of up to $\SI{8.0}{\electronvolt}$ reported for oxide defects in previous works~\cite{rzepa2018comphy}. In order to obtain a unique, physical parameter set, we therefore recommend to restrict the simulations to the so-called \textit{linear coupling regime} defined by $R=1$~\cite{nitzan2006chemical, PhysRevB.20.5084}. This choice is supported by extensive theoretical DFT studies on oxide defects in a-SiO$_2$~\cite{goes2018}. Besides this empirical justification, one could also make an argument based on the curvatures of the PECs being related to the phonon frequencies in the material. Since in an amorphous material many different phonon modes will be involved in a charge transition, the phonon frequency changes in different charge states will mostly average out, resulting in an effective curvature ratio of $R=1$.

Although we demonstrated the $(R,E_\mathrm{R})$ correlation in the 2-state NMP model, we cannot rigorously rule out the existence of other cross-correlations at the current time. For example, we suspect a similar correlation between $\Delta Q$ and $E_\mathrm{R}$ in the quantum mechanical model, however, this is subject to further investigations.

\begin{figure*}[t]
	\centering
	\includegraphics[width=2\columnwidth]{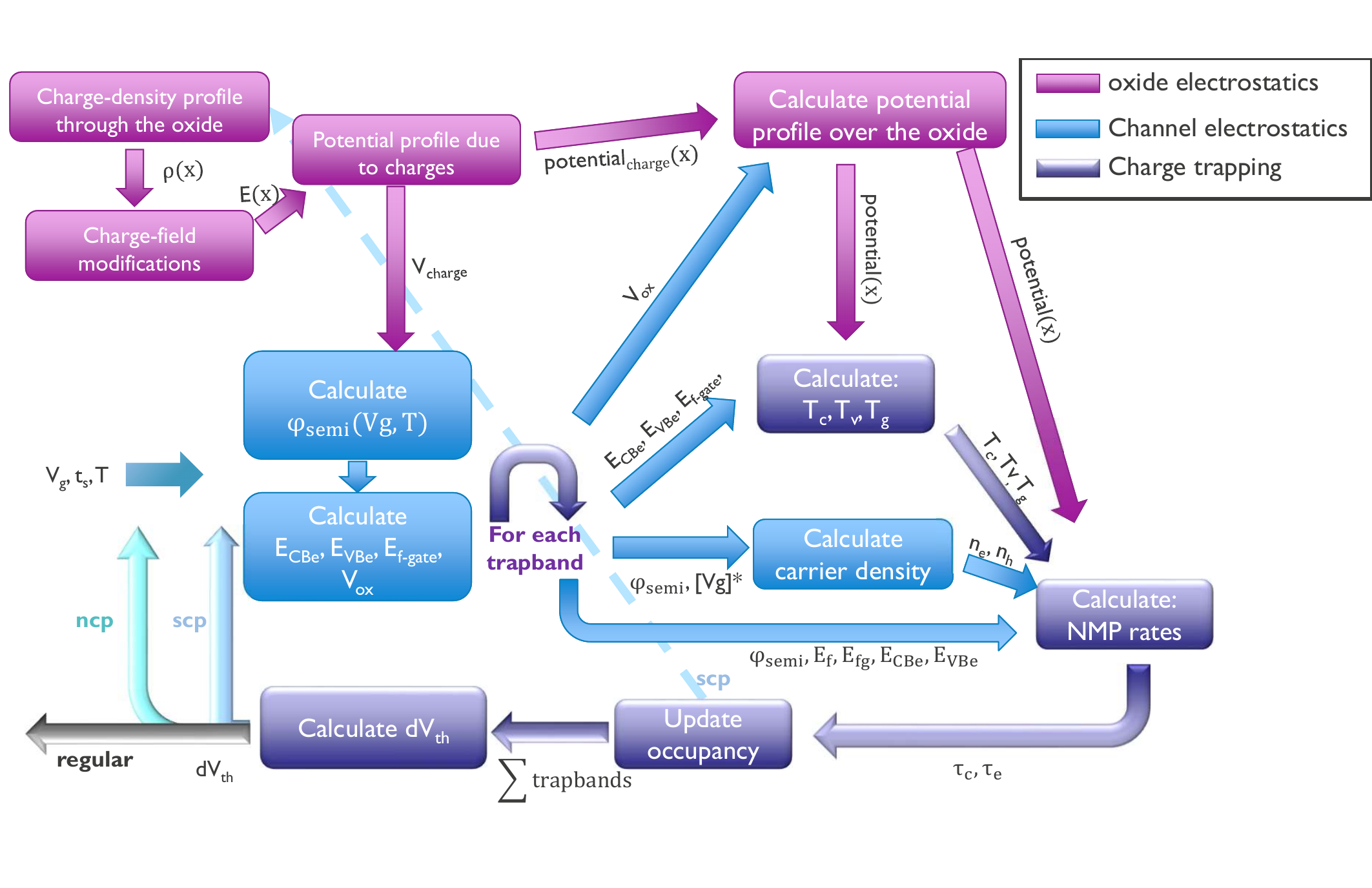}
	\caption{A schematic representation of the internal structure of Comphy and the flow of information for calculating a $\Delta V_\mathrm{th}$-response (bottom left) starting from a set of input quantities (middle left). Boxes correspond to different stages in the calculation (each focused on obtaining a specific set of intermediary quantities), colors correspond to larger sections (stages grouped by a common denominator), arrows correspond to quantities being transferred from one stage to another. ``Regular'', ``ncp'' and ``scp'' refer to different ways of coupling charged traps to the electric field. ``Regular'' implies no coupling, ``ncp'' implies the charges of the current field will influence the next time step, ``scp'' implies the charges are coupled self-consistently and for every time step a convergent solution is acquired.} 
	\label{fig:ComphyStructure}\vspace*{-5mm}
\end{figure*}

\section{Code Structure}

Figure \ref{fig:ComphyStructure} illustrates the internal logic of the simulation engine in \textit{Comphy}. Each block represents a set of operations (functions) with the arrows illustrating the data propagation between the different blocks. This flow chart assumes all device parameters have already been loaded. The simulation starts at the left side where the gate bias $V_\mathrm{G}$ as a function of the time $t$ at a certain temperature $T$ is given as input quantity and is propagated through the framework, eventually producing a $\Delta V_{th}$ prediction (black arrow bottom left). Propagating through the framework, operations are subdivided into three major categories depending on whether the calculated quantity belongs to the channel (blue), the dielectric (purple) or the charge trapping at the defects (violet). Each block calculates one or more quantities using the provided input from earlier blocks up the chain. The flow is not static and depends on the features used/physics requested for a simulation. To improve the overall performance and to reduce redundant calculations, transparent caching is applied at various stages, as e.g. the electrostatic profiles may be requested multiple times by different blocks. In addition, certain quantities remain identical (given the same input) throughout the waveform, especially for cases where the coupling between the electric field profile and the trapped charges is neglected.\newline

Elaborating on this coupling, the black, teal and cyan arrows in the bottom left corner of Figure \ref{fig:2state_NMP} represent the various available charge-field coupling schemes in \textit{Comphy} (``regular'', ``ncp'' and ``scp'' respectively). These allow the user various levels of speed-accuracy trade-offs by either not coupling the trapped charge and the electrostatics (“regular”), accounting for
the trapped charge using an equivalent projected charge for the next timepoint (non-self consistent coupling, “ncp”) or coupling the trapped charge with the oxide field and iteratively solving for a consistent solution (self-consistent coupling, “scp”).

\section{Parameter Extraction Methods}
\label{sec:parameter_extraction}
One of the main goals in device reliability physics is to accurately predict the performance and degradation of a device over its lifetime. Since the degradation processes happen on the timescale of years under normal operating conditions, accelerated stress conditions are used to observe the degradation in a reasonable time frame of hours to days. The main difficulty is to extrapolate the observations back to operating conditions in a physically meaningful way. Due to their simplicity, empirical power laws are frequently used in industry to assess the device reliability, however their predictive capabilities are limited due to the lack of a physical model. Instead, \textit{Comphy} relies on a physical description of charge trapping based on the rigorous NMP framework. This approach does not only provide a more adequate extrapolation scheme, but also allows to compare the defect parameters of the simulation to theoretical DFT studies in order to identify possible defect candidates responsible for the observed degradation.

\subsection{Fitting Problem}
\begin{figure}[t]
	\centering
	\includegraphics[width=\columnwidth]{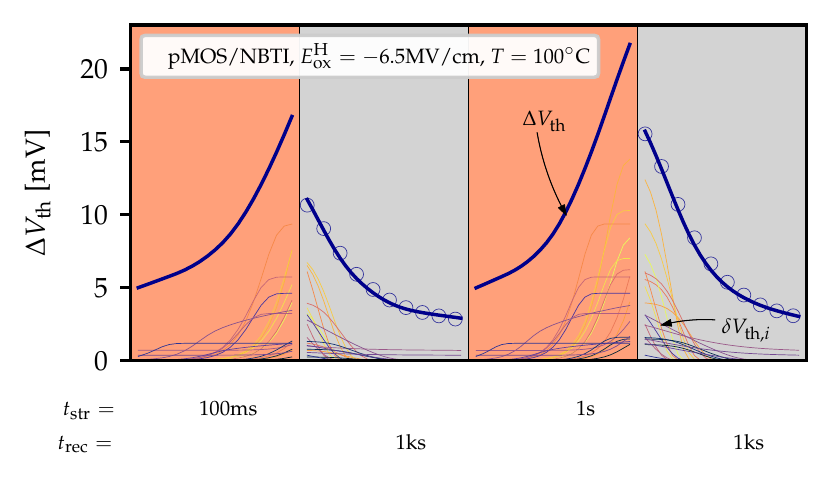}
	\caption{Assuming non-interacting defects, the transient $\Delta V_\mathrm{th}$ during an eMSM sequence can be linearly decomposed into small contributions $\delta V_\mathrm{th,i}$ from individual defects. The red and gray regions indicate the stress and recovery phases of the eMSM sequence respectively.}
	\label{fig:superposition}
\end{figure}
Due to the amorphous nature of the gate dielectric, the defect parameters are usually distributed, resulting in capture and emission time constants spanning many orders of magnitude. Furthermore, most experiments on $\Delta V_\mathrm{th}$ degradation are carried out on large-area devices, where only the collective response of a whole defect ensemble is observable. Typically, eMSM schemes~\cite{kaczer2008} are used in experiments, resulting in recovery curves as depicted in \reffig{fig:superposition}. Using the 2-state NMP model, one can deduce from \refeq{eq:2state} that these recovery traces have to be composed of multiple exponentially decaying functions. 
Assuming non-interacting defects, the total $\Delta V_\mathrm{th}$ can be expressed as a linear superposition of individual defect contributions, i.e.
\begin{equation}\label{eq:superpos}
\Delta V_\mathrm{th}(t,V_\mathrm{G},T)=\int_\Omega N(\boldsymbol{p})\cdot \delta V_\mathrm{th}(t,V_\mathrm{G},T;\boldsymbol{p}) \mathrm{d}\boldsymbol{p}\,.
\end{equation}
Here, $\boldsymbol{p}$ is a set of defect parameters similar to \refeq{eq:tupel}, $\delta V_\mathrm{th}$ is the contribution of a single defect with parameters $\boldsymbol{p}$ and $N(\boldsymbol{p})$ is the distribution function in the parameter space $\Omega$.

Hence, extracting defect parameters from experiments is equivalent to finding a suitable distribution function $N(\boldsymbol{p})$ in order to fulfill \refeq{eq:superpos}. However, such an \textit{inverse problem}~\cite{tarantola2005inverse} is mathematically ill-posed and requires to impose further restrictions on $N(\boldsymbol{p})$ to obtain a physical solution. Another relevant example of such a problem would be the \textit{multiexponential analysis} required in deep level transient spectroscopy (DLTS)~\cite{doi:10.1063/1.344973, istratov1999exponential}.

In the following, we discuss two different approaches implemented in Comphy to deal with this parameter extraction problem.

\subsection{Gaussian Defect Bands}
The main challenge in solving \refeq{eq:superpos} is that the unknown $N(\boldsymbol{p})$ is a scalar field and hence has far too many degrees of freedom. A straightforward approach to obtain a solution is to enforce a certain distribution shape. The most common choice found in literature is a normal distribution for the energetic parameters $E_\mathrm{T}$ and $E_\mathrm{R}$ resulting in a \textit{Gaussian defect band}~\cite{rzepa2018comphy}. The spatial variable $x_\mathrm{T}$, on the other hand, is usually assumed to be uniform, leading to a constant defect density in a certain region of the oxide. Besides this simple model, an exponentially decaying spatial distribution is often chosen to reflect the increasing defect density towards the interface~\cite{franco2021}.
\begin{figure}[t]
	\centering
	\includegraphics[width=\columnwidth]{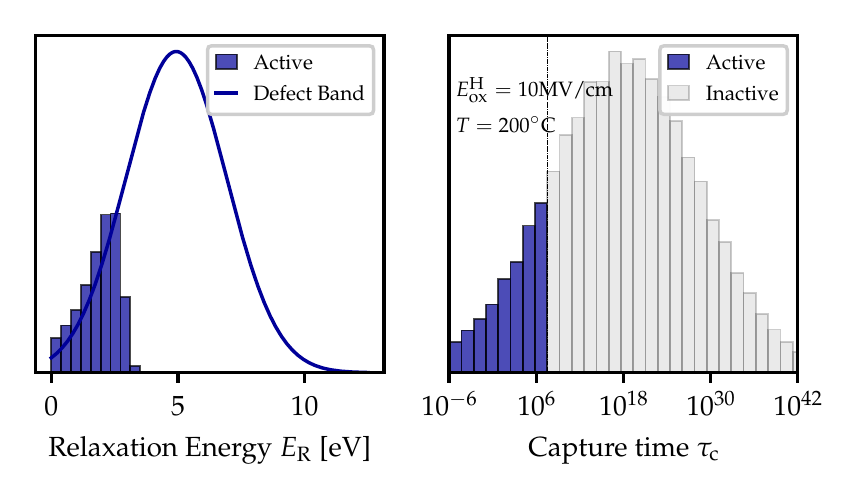}
	\caption{Extracted Gaussian trap band for electron traps in SiO$_2$~\cite{rzepa2018comphy} \textbf{(left)}. However, most defects in the band have exceedingly large capture time constants, rendering them electrically inactive. Only a small portion at the distribution tail can become charged within the measurement window \textbf{(right)}.}
	\label{fig:gauss}
\end{figure}
By fixing the mathematical form of the distribution, only its defining parameters, like the mean and standard deviation in the case of Gaussian defect bands, have to be optimized to fit the experimental $\Delta V_\mathrm{th}$. This can be achieved by using the simplex-method or other local optimizers.

The concept of Gaussian defect bands is intuitive and has been used successfully to reproduce experimental data across a large variety of different devices~\cite{rzepa2018comphy,sullivan2020, rzepa2019, claes2020}. However, artificially constraining the shape of the parameter distribution might lead to some artifacts making it harder to provide physical interpretations and to link the obtained distributions to theoretical DFT studies. One example for such a misrepresentation is shown in \reffig{fig:gauss} for the case of electron traps in SiO$_2$\,\cite{rzepa2018comphy}. While the extracted defect band would suggest a mean relaxation energy of $\SI{5}{\electronvolt}$, closer examination shows that even under severe stress conditions most defects within this band are electrically inactive due to their exceedingly large time constants. Instead, when filtering out too large capture times, only the tails of the Gaussian band contribute to the observed charge trapping, leading to a significantly altered parameter distribution. For this reason, it is advisable to check the corresponding time constants when using Gaussian bands and to truncate the distributions if appropriate.

Another drawback of Gaussian bands is that the optimization procedure often needs manual intervention and a good initial guess. Especially in SiC based devices, where multiple different defect bands are involved as a result of the increased stoichiometric complexity near the interface~\cite{afanasev1997}, the optimization becomes increasingly difficult and time consuming.

\subsection{Effective Single Defect Decomposition (ESiD)}\label{sec:esid}
In order to eliminate the drawbacks of predefined distribution functions, we developed a novel extraction method named \textit{Effective Single Defect Decomposition} (ESiD)~\cite{waldhoer2021} which is capable of inferring an unrestricted physical defect distribution from experimental eMSM sequences in a semiautomatic way.

\subsubsection{Algorithm}
Going back to the original problem \refeq{eq:superpos}, we directly exploit the fact that the macroscopic $\Delta V_\mathrm{th}$ is a linear superposition of independent defect responses denoted by $\delta V_\mathrm{th}$. While this assumption is strictly true only in the limit of low defect concentrations, it is still a good approximation for typically observed defect concentrations in the range of $\SI{e18}{}-\SI{e20}{\per\cubic\centi\meter}$. Regardless, the quality of this approximation can always be checked retrospectively by using the resulting defect distributions in a self-consistent Poisson (scp) calculation and comparison to the non-selfconsistent results used for obtaining aforementioned distributions in the first place. Under this assumption, finding an optimal distribution function $N(\boldsymbol{p})$ can be recast into the non-negative linear least square (NNLS) problem
\begin{equation}\label{eq:unreg}
N(\boldsymbol{p})=\underset{\hat{N}\geq 0}{\mathrm{arg\,min}} \left|\Delta V_\mathrm{th}-\int_\Omega \hat{N}(\boldsymbol{p})\delta V_\mathrm{th} \mathrm{d}\boldsymbol{p}\,\right|^2\,.
\end{equation}
Note that the non-negativity constraint is essential here, since negative values for the defect density would constitute an unphysical solution. However, \refeq{eq:unreg} is merely a reformulation of the original problem and hence is still ill-posed. 

\begin{figure}[t]
	\centering
	\includegraphics[width=\columnwidth]{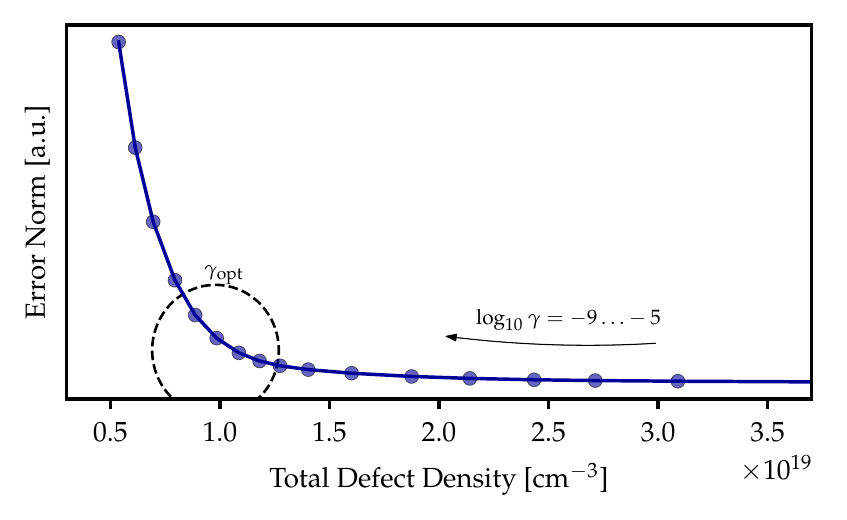}
	\caption{Impact of the regularization term $\gamma$ on the obtained solutions. The optimal value for $\gamma$ is obtained at the ``corner'' of the L-shaped curve, before a further increase of $\gamma$ leads to a steep increase in the error norm of the solution.}
	\label{fig:error_reg}
\end{figure}

Naively trying to solve the optimization problem as stated above would lead to unstable solutions which are highly sensitive to noise in the input data. In order to obtain a physically meaningful solution, the problem has to be regularized. The implementation in Comphy uses the \textit{Tikhonov} scheme~\cite{tikhonov1995numerical}, in which a regularization term depending on the total density
\begin{equation}
\hat{N}_\mathrm{tot}=\int_\Omega \hat{N}(\boldsymbol{p})\mathrm{d}\boldsymbol{p}
\end{equation}
is introduced. In this scheme, the term to be minimized is given by
\begin{equation}
 \left|\Delta V_\mathrm{th}-\int_\Omega \hat{N}(\boldsymbol{p})\delta V_\mathrm{th} \mathrm{d}\boldsymbol{p}\,\right|^2+\gamma^2 N_\mathrm{tot}^2
\end{equation}
with $\gamma$ being an adjustable regularization parameter. The effects of $\gamma$ on the optimization are twofold. It favors solutions which require only a low defect density to match the experimental data and, although not obvious, also implicitly forces the solutions to be smooth. This can be shown by performing a singular-value decomposition (SVD) on the linear optimization problem and realizing that $\gamma$ sets a lower bound for the smallest possible singular values, which are responsible for discontinuous solutions. In practice, $\gamma$ has to be adjusted to the particular problem at hand. However, as illustrated in \reffig{fig:error_reg} a reasonable value can be estimated by plotting the total defect density versus the error norm to the experimental data for different values of $\gamma$, resulting in a L-shaped curve. As can be seen, for too small values, the approximation error is very low at the expense of a very high defect concentration. Quite to the contrary, if $\gamma$ is too large, the problem becomes overregularized, meaning that there is a steep increase in approximation error when further increasing $\gamma$. According to the \textit{L-criterion}~\cite{CALVETTI2000423}, the optimal value for $\gamma$ lies at the ``corner'' of this L-shaped curve, providing a good compromise between accurate representation of the experimental data with a reasonably small defect concentration.

\subsubsection{Application}
For an efficient implementation in \textit{Comphy}, we only employ ESiD for the energetic parameters $E_\mathrm{T}$ and $E_\mathrm{R}$ since they are most relevant for identifying possible defect candidates. As mentioned earlier, we fix $R=1$ in order to eliminate cross-correlations between the model parameters. The distributions along $x_\mathrm{T}$ and $\Delta Q$ are assumed to be uniform, meaning that the responses along these dimensions can be added up before optimization with ESiD. By doing so, we can reduce the parameter extraction to determining a 2-dimensional $(E_\mathrm{T}, E_\mathrm{R})$ distribution function. The NNLS optimization problem can then be easily discretized by defining a search region $[E_\mathrm{T,min}, E_\mathrm{T,max}]\times [E_\mathrm{R,min}, E_\mathrm{R,max}]$ on a uniform grid. 

\begin{figure}[t]
	\centering
	\includegraphics[width=\columnwidth]{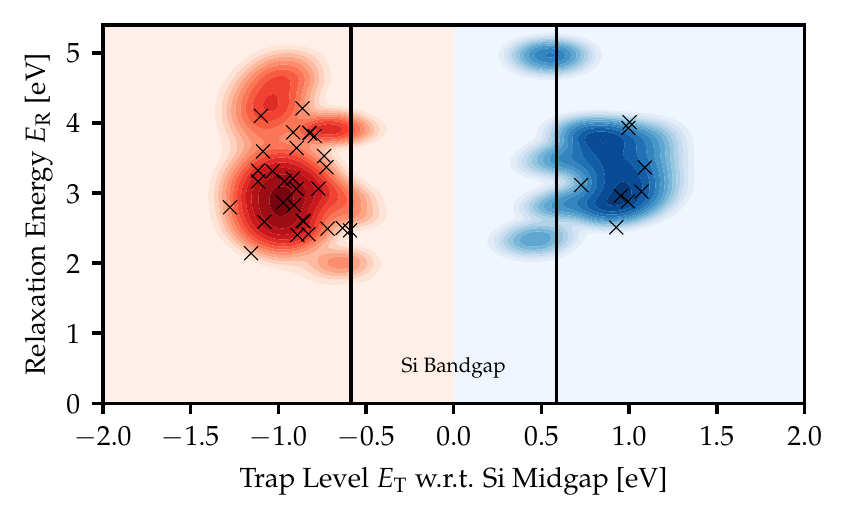}
	\caption{$(E_\mathrm{T}, E_\mathrm{R})$ maps extracted from large-area SiON devices using ESiD~\cite{waldhoer2021} (heatmap). The distributions agree well with individual electron and hole traps (crosses) extracted with TDDS from similar small-area devices~\cite{grasser2014, WALTL2020113746}.}
	\label{fig:et_er_map}
\end{figure}

\begin{figure}[t]
	\centering
	\includegraphics[width=\columnwidth]{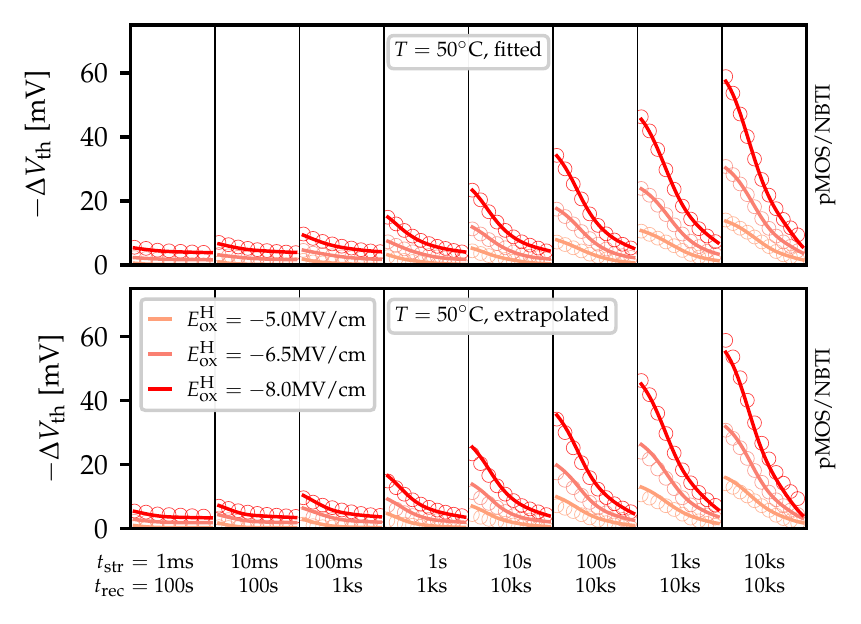}
	\caption{\textbf{Top:} eMSM traces for SiON device~\cite{waldhoer2021} at T=$\SI{50}{\celsius}$ fitted with ESiD. \textbf{Bottom:} A different ESiD model was calibrated to experimental data obtained at T=$\SI{100}{\celsius}$ and T=$\SI{150}{\celsius}$ and subsequently extrapolated down to T=$\SI{50}{\celsius}$. As can be seen, the prediction of this model for the $\Delta V_\mathrm{th}$ degradation shows excellent agreement with the actual measurement data.}
	\label{fig:fit_cal}
\end{figure}

An example of such an extracted $(E_\mathrm{T}, E_\mathrm{R})$ map is shown in \reffig{fig:et_er_map} for electron and hole traps in SiON dielectrics~\cite{waldhoer2021}. Note that this map extracted from large-area devices with eMSM agrees well with data obtained for individual defects by means of TDDS on similar small-area devices~\cite{grasser2014,WALTL2020113746}. Although both methods give comparable results, the ESiD distribution was extracted automatically from measurement data obtained in a matter of days, whereas the single-defect characterization took several months. This clearly demonstrates the advantages provided by the ESiD extraction scheme. To further show the predictive capabilities of a device model created automatically by ESiD from eMSM data, \reffig{fig:fit_cal} shows a comparison between a model which was actually fitted to traces at $T=\SI{50}{\celsius}$ (top) and a different model which was extracted from experimental data at $T=\SI{100}{\celsius}$ and $\SI{150}{\celsius}$ and subsequently extrapolated down to $\SI{50}{\celsius}$ (bottom). As can be seen, the extrapolated degradation from the device model almost perfectly matches the actual measurement data. Note that in order to obtain a good model, the experimental data used for parameter extraction needs to include high temperature traces in order to also probe slower defects responsible for the long-term degradation at lower temperatures.

\section{Gate-leakage Currents}
\begin{figure}[t]
	\centering
	\includegraphics[width=\columnwidth]{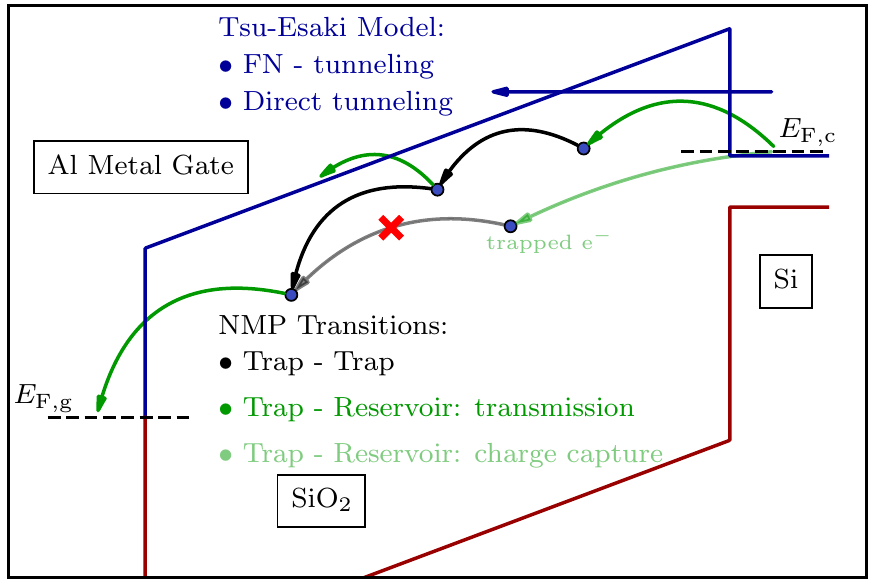}
	\caption{The different tunneling mechanisms that typically appear in a MOS gate stack are shown schematically. The band to band tunneling mechanisms (green) are included in the well-known Tsu-Esaki model as implemented in Comphy, while all trap-assisted tunneling reactions (blue) are described by two-state NMP transitions. }
	\label{fig:tunneling}
\end{figure}

Besides charge-trapping which leads to considerable $V_\mathrm{th}$ shifts in MOSFETs, as described in the previous sections, time-zero gate-leakage currents pose a severe threat for power dissipation and to gate oxide reliability. Within \Comphy, an efficient modeling approach is incorporated to enable the computation of these leakage currents that arise from the same inelastic charge transfer reactions driving BTI, but without additional parameters compared to the BTI simulations~\cite{schleich2022singleI}. Depending on the gate stack material and defect properties within the insulating layers, either band-to-band tunneling or defects acting as charge transition centers between gate and channel dominate the leakage mechanism. As shown in \reffig{fig:tunneling}, the trap-assisted tunneling (TAT) component appears as either a single- or multi-step process. 

\subsection{Tsu-Esaki Model}\label{sec:tsu_esaki}

The most commonly used approximation of the current density due to tunneling of charge carriers between two carrier reservoirs separated by an energetic barrier is given by the Tsu-Esaki formalism~\cite{tsu1973tunneling}.  For electron tunneling the current density is given as
\begin{align}
J_{\mathrm{TE,e}} &= \frac{4\pi m_\mathrm{e} q_0}{h^3} \int_{E_\text{CB}}^{\infty} \vartheta_\mathrm{WKB}\left(E\right) N_\mathrm{e}\left(E\right) \mathrm{d}E
\label{equ:tsu_esaki_currdens}
\end{align}
with the effective electron mass $m_e$ in the semiconductor channel and the supply function
\begin{align}
N_e \left(E\right) = k_\text{B}T \, \text{ln} \left( \frac{1+\text{exp}\left(-\frac{E-E_\text{F1}}{k_\text{B}T}\right)}{1+\text{exp}\left(-\frac{E-E_\text{F2}}{k_\text{B}T}\right)} \right).
\label{equ:supply_func}
\end{align}
accounting for the  Fermi-dirac distribution of the carriers in both electrodes. The tunneling probability is calculated using a WKB approximation by
\begin{align}
\vartheta_\mathrm{WKB}\left(E\right) = \text{exp} \left( -\frac{4\pi}{h} \int_{x_1}^{x_2} \sqrt{2m_{e,\mathrm{diel}}\left( W\left( x \right) - E \right)} \, \dd x \right)
\label{equ:WKB_full}
\end{align}with the tunneling effective mass in the dielectric layer $m_{e,\mathrm{diel}}$ and the energetic barrier $W\left( x \right)$. Depending on the shape of $W\left( x \right)$, i.e. triangular or trapezoidal, \refeq{equ:tsu_esaki_currdens} evolves to the well-known Fowler-Nordheim or direct tunneling formulas.

\subsection{Efficient Trap-Assisted Tunneling Current Computation}\label{sec:tat}
\begin{figure}[t!]
    \centering
    \includegraphics[width=\columnwidth]{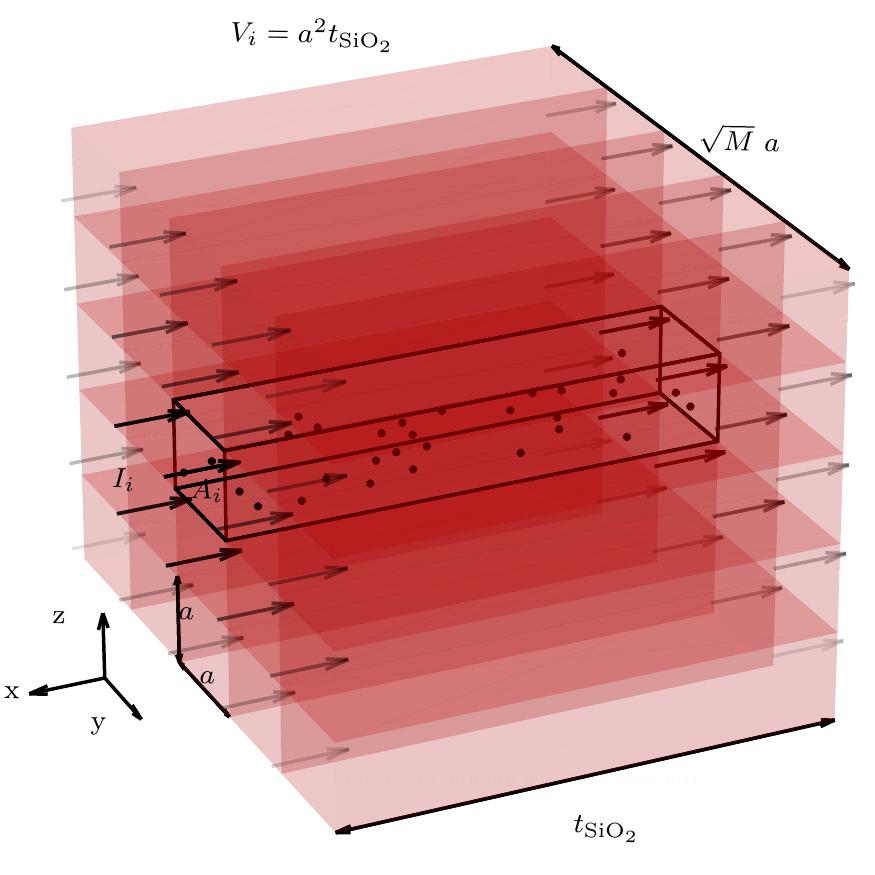}
    \caption{The three dimensional spatial simulation domain is divided into $M$ slabs in the dielectric layer. Within each slab a Poisson distributed number of defects $N$ with mean value $\overline{N}$ is sampled throughout the volume $V_i = a^2 t_{\mathrm{SiO}_2}$, with $a$ resulting from the defect density $N_\mathrm{T}$.}
    \label{fig:slabs}
\end{figure}
Contrary to charge capture at oxide defects in large-area devices, the calculation of charge transfer between individual defects in amorphous materials is not feasible with a simulation on grids with equidistant grid-points. 
Therefore, for the calculation of the TAT current within an amorphous dielectric, the defects are randomly sampled based on a uniform distribution in a three dimensional volume $V_i$, as schematically shown in \reffig{fig:slabs}. The defect density $N_\mathrm{T}$ and average defect number $\overline{N}$ thereby define $V_i = \overline{N} / N_\mathrm{T}$. The actual defect number $N$ is drawn from a Poisson distribution within each simulated $V_i$. From the three dimensional sampling, the distances $d_\mathrm{T}$ between the defects can be calculated, while electrostatic quantities are computed in a one-dimensional simulation space, as described in~\cite{rzepa2018comphy}. Defects sampled with unphysically low distances to each other (below $\SI{1}{\nano\meter}$) are removed from the sample and redrawn. In order to capture the stochastic variation of the simulation results, the simulation can be performed within $M$ simulation volumes in parallel. Note that the ESiD extraction scheme is not applicable to obtain the model parameters from experiments due to the nonlinearity caused by the defect-defect interaction.  Here, a  Gaussian distribution is used for the sampling of the energetic defect parameters $E_\mathrm{T}$ and $E_\mathrm{R}$ which fully determine the classically approximated PECs of the two charge states of a defect within the NMP model. This representation allows for the calculation of the energetic barriers $\varepsilon_{ij}$ that determine the classical charge transition rates $k$.  As described in detail in \cite{schleich2022singleI}, considering all possible charge transfer paths, the trap-assisted leakage current can be computed based on a generalized Shockley-Ramo theorem~\cite{jungemann2014dc} by
\begin{multline}
I_{\mathrm{G,TAT}} = \underbrace{C_\mathrm{ox} \derivative{V_\mathrm{G}}{t}}_{\text{displacement \linebreak current}} \\ + \underbrace{q_0 \sum_i^N k_{\mathrm{e},i,\mathrm{gate}} f_i - k_{\mathrm{c},i,\mathrm{gate}} \left(1-f_i\right)}_{\text{single-TAT current}} \\ 
+ \underbrace{q_0 \sum_{i}^{N}  \sum_{j\neq i}^{N} k_{\mathrm{e},ij} f_i \left( 1 - f_j \right) \frac{x_{i}-x_{j}}{t_\mathrm{ox}}}_{\text{multi-TAT current}} \\ 
+ \underbrace{q_0 \sum_i^N \left[ k_{\mathrm{c},i,\mathrm{channel}} \left(1 - f_i\right) - k_{\mathrm{e},i,\mathrm{channel}} f_i \right] \big( 1 - \frac{x_i}{t_\mathrm{ox}} \big)}_{\text{charge trapping current (channel)}}.
\label{equ:TAT_cur}
\end{multline}
Hereby the required NMP parameters to compute the charge transition rates $k_{ij}$ for the multi-TAT current can be readily derived from the PECs for the carrier reservoir to defect charge transfer reactions. This parameter conversion is performed internally in \Comphy{} and valid within the limitations presented in \cite{schleich2022singleI}. Hence no additional simulation parameters are required for including the multi-TAT current computation. Due to the coupling of the defects in multi-TAT mode, the arising non-linear coupled system of Master equations to compute the defect occupations needs to be solved. The Newton scheme used to calculate the occupations in multi-TAT mode leads to a reduced computational efficiency compared to single-TAT mode, whose performance is comparable to the simple \dVth{} calculation. In addition, the single-TAT mode is also accessible by ESiD. As shown in \cite{schleich2022singleI} for many applications a multi-step TAT current only becomes relevant at large $N_\mathrm{T}$ and for low relaxation energies $E_\mathrm{R}$ of the conducting defect bands. Due to these prerequisites, single-step TAT already provides efficient and accurate results for the majority of technologies.

\subsection{Application Example: Leakage currents in high-$\kappa$ capacitor}

\begin{figure}[t]
	\centering
	\includegraphics[width=\columnwidth]{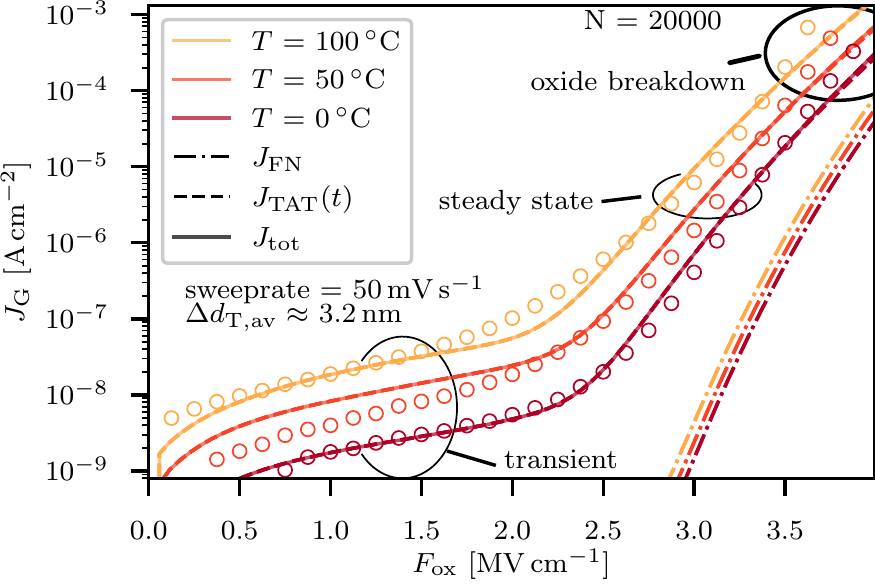}
	\caption{Gate leakage currents measured (circles) on a TZT MIM capacitor show two branches of oxide field dependence. The simulation (lines) reveals that the shallow branch at low $F_\mathrm{ox}$ can be explained by a transient charge trapping current. The steep branch at increased $F_\mathrm{ox}$, on the other hand, is explained by fast defect transitions between the electrodes via the oxide conduction band.}
	\label{fig:zro_cap}
\end{figure}

Storage capacitors for RAM application fabricated from a thin ZrO$_\mathrm{2}$ layer stacked between two TiN electrodes (TZT) enable further scaling of the memory cells. However, these capacitors show increased thermally activated leakage currents at low to medium electric field strengths~\cite{jegert2011ultimate, padovani2019}. By using the Comphy framework extension for leakage current computation as described in the previous section, the measured leakage currents are reproduced accurately for a capacitor employing a \SI{8}{\nano\meter} thick ZrO$_\mathrm{2}$ layer, as shown in \reffig{fig:zro_cap}. Therefore, two defect bands are required to explain the two branches with a shallow and steep current increase over the applied gate bias. These defect bands exhibit relatively large relaxation energies $E_\mathrm{R}$ of about \SI{2.6}{\electronvolt} for the charge trapping band on the one hand and low $E_\mathrm{R}$ of about \SI{0.8}{\electronvolt} for the defects responsible for the steady-state TAT current. These low $E_\mathrm{R}$ show excellent parameter agreement with those calculated with DFT for a statistically relevant number of  model structures with polarons in partly recrystallized ZrO$_\mathrm{2}$~\cite{schleich2022singleII}.  

\section{Cryogenic Charge Trapping}\label{sec:cryo}
\begin{figure}[t]
    \centering
    \includegraphics[width=\columnwidth]{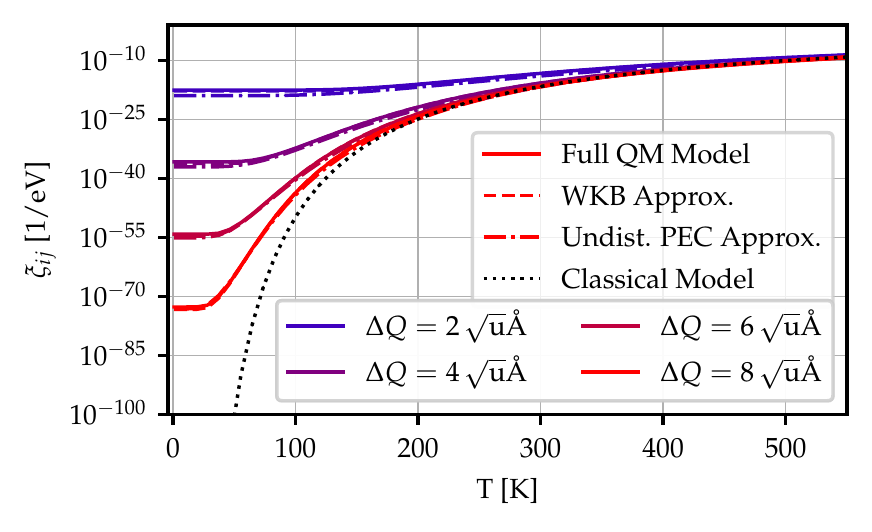}
    \caption{The lineshape function $\zeta_{ij}$ for various configuration coordinate offsets \dq~for $\Er=\SI{2.5}{\eV}$, $R=1$ and $\dE=\SI{0}{\eV}$ computed with the full quantum mechanical model (solid), the WKB-based model (dashed) and the undistorted PEC model (dash-dotted) becomes temperature independent towards cryogenic temperatures while the classical model (dotted) freezes out completely.}
    \label{fig:cryo_rates}
\end{figure}

\begin{figure}[t]
	\centering
	\includegraphics[width=\columnwidth]{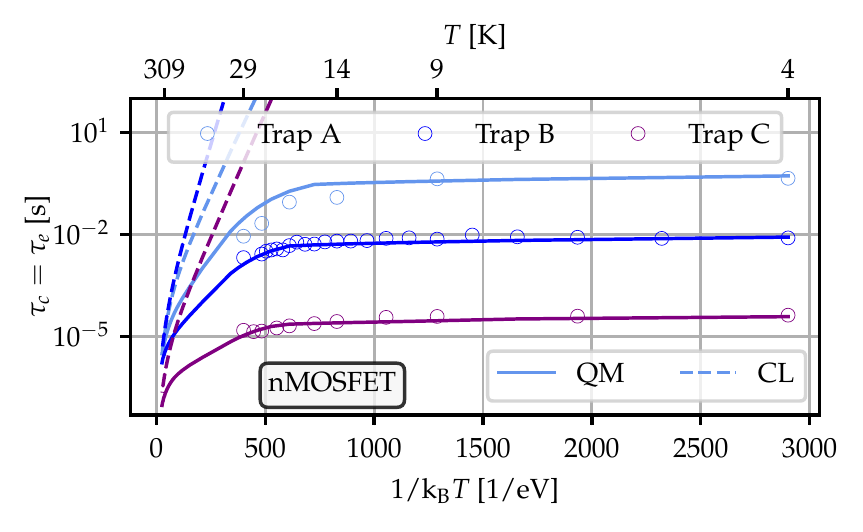}
	\caption{Experimental time constants observed in RTN signals at cryogenic temperatures together with fits of the classical and full quantum mechanical model~\cite{Michl2021EvidenceOfTunneling}. As can be seen, the temperature independence of the time constants cannot be described within the classical model.}
	\label{fig:cryo_rtn}
\end{figure}

The classical approximation of the 2-state NMP transition rates~\refeq{eq:NMPrates_2state_cl} is well suited for describing charge transfer kinetics above room temperature, however, it breaks down at cryogenic temperatures. While multiple studies show active charge trapping at cryogenic temperatures~\cite{Michl2021EfficientModelingOf2, Knobloch2020AnalysisOfSingle, Michl2021EvidenceOfTunneling,  Grill2020ReliabilityAndVariability} the classical limit
\begin{align}
    \lim_{T\to\SI{0}{\kelvin}} k_{ij}^\mathrm{cl}(T)=0
\end{align}
predicts a total freeze out of charge transitions between neutral and charged defect configurations. Thus, it is necessary to use the full quantum mechanical version of the 2-state NMP model for cryogenic modeling. Here, the transition rates at cryogenic temperatures are not dominated by the height of the classical barrier ~$\varepsilon_{ij}$ but by the overlap of the vibrational wave functions as can be seen in \reffig{fig:2state_NMP}. Towards cryogenic temperatures, the transition rate is dominated by the overlaps of the vibrational ground state, because the low thermal energy restricts access to excited states. As the most obvious consequence, the lineshape function which is proportional to the transition rate becomes temperature independent as shown in \reffig{fig:cryo_rates}. This transition from one atomic configuration to another at energies below the classical barrier is called nuclear tunneling. This temperature independence is also demonstrated  for defects causing RTN at cryogenic temperatures in \reffig{fig:cryo_rtn}. The computation of the full quantum mechanical transition rate is numerically expensive, because it requires to compute the eigenstates of the potential energy surfaces and the corresponding overlap functions which then need to be summed up. This procedure is not suitable for reliability simulations with thousands of defects. Therefore, a numerically effective model based on a WKB approximation of the vibrational wave functions was developed in~\cite{Michl2021EfficientModelingOf1} and implemented in  \textit{Comphy v3.0}.

\subsection{WKB-based approximation}
\begin{figure}[t]
	\centering
	\includegraphics[width=\columnwidth]{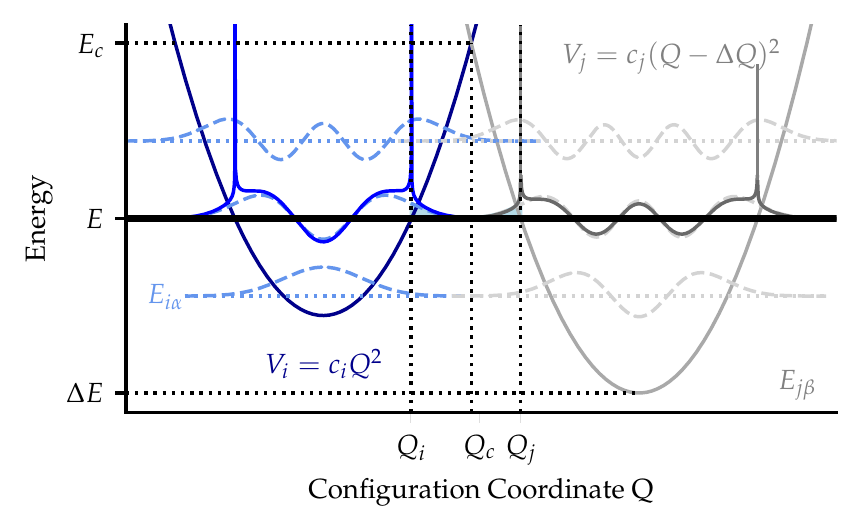}
	\caption{The vibrational wave functions (dashed) of the potential energy curves representing the neutral and charged states can be approximated using WKB-based wave functions (solid). The overlap of the wavefunctions is dominated by their exponentially decaying parts which allows the formulation of a simplified expression for the charge transition rates.}
	\label{fig:WKB_approx}
\end{figure}

Within the developed approximation scheme for the 2-state NMP model, the exact vibrational wave functions are replaced by their respective WKB-approximations as shown in \reffig{fig:WKB_approx}. 

At cryogenic temperatures, the overlaps of the wave functions which are needed for the computation of~\refeq{} are dominated by the exponentially decaying part of the WKB wave function. This allows to simplify the overlap integral \refeq{eq:fc_lsf} to a continuous form and approximate it analytically using the stationary phase method~\cite{markvart1981}. The transition rate $k_{ij}$ then evaluates to
\begin{align}
    k_{ij}(T) = C(E^*)\exp\left(\frac{-E^*}{k_\mathrm{B}T}+\varphi(E^*)\right)\sqrt{\frac{2\pi}{\varphi''(E^*)}}\,,
\end{align}
where $\varphi(E)$ can be expressed analytically as shown in \cite{Michl2021EfficientModelingOf1}. The energy $E^*$ can be obtained via the saddle-point method using a numerical optimizer for solving
\begin{align}
    \label{eq:cryo_opt}
    \frac{\mathrm{d}\varphi(E)}{\mathrm{d}E}\bigg\rvert_{E=E^*}=\frac{1}{k_\mathrm{B}T}.
\end{align}
This model gives very similar results to the full quantum mechanical (FQM) model while also being computationally much more efficient~\cite{Michl2021EfficientModelingOf1}. This is shown exemplary in \reffig{fig:cryo_rates}, where the lineshape function of the WKB-based approximation and the (FQM) model are shown for $\Er=\SI{2.5}{\eV}$, $R=1$ and $\dE=\SI{0}{\eV}$ for a wide range of temperatures and configuration coordinate offsets \dq. This example shows the ability of describing the non-freeze-out behavior of the quantum mechanical transition rates compared to the classical model and the experimentally observed temperature independent behavior. 

However, due to the required optimization step~\refeq{eq:cryo_opt} no closed form expression is available which results in a considerably slower computation compared to the classical approximation. For further numerical improvement a model based on the assumption of linear coupling ($R=1$) is developed.

\subsection{Closed Form for Linear Coupling}
For $R=1$, the curvatures of the PECs do not change upon charge trapping, implying that the effective phonon frequencies are also the same in both defect charge states ($\omega_i=\omega_j=\omega$). In this case, it is possible to develop a closed form expression for the quantum mechanical transition rates in the strong electron-phonon coupling regime at cryogenic temperatures~\cite{Freed1970MultiphononProcessesIn}. Here, the transition rate $k_{ij}$ can be computed as 
\begin{align}
\begin{split}
    k_{ij}(T) = \omega \frac{\sqrt{2\pi}}{2D\hbar}\big((1+\coth \alpha)\,\mathrm{e}^{{-E_\mathrm{-}}/{2D^2\hbar^2}}-\\
    (1-\coth \alpha)\,\mathrm{e}^{{-E_\mathrm{+}}/{2D^2\hbar^2}} \big)
\end{split}
\end{align}
with 
$D=\sqrt{\omega\Er (2n+1)/\hbar}$, $n=1/(\exp(\hbar \omega/k_\mathrm{B}T)-1)$, $\alpha=\hbar\omega/2k_\mathrm{B}T$ and $E_{\pm} = \dE \pm \hbar\omega -\Er$. The phonon frequency $\omega$ can be obtained for a given configuration coordinate offset \dq~from the relaxation energy by
\begin{align}
    \omega=\frac{\sqrt{2\Er}}{\dq}.
\end{align}
As shown in the example in \reffig{fig:cryo_rates}, the closed form expression for linear coupling gives the same results as the FQM model over a wide temperature range.  Furthermore, computing this analytical expression is comparable to the classical model in terms of computational costs and is thus well suited for efficient reliability simulations with thousands of defects in a device.

\begin{figure}[t]
	\centering
	\includegraphics[width=\columnwidth]{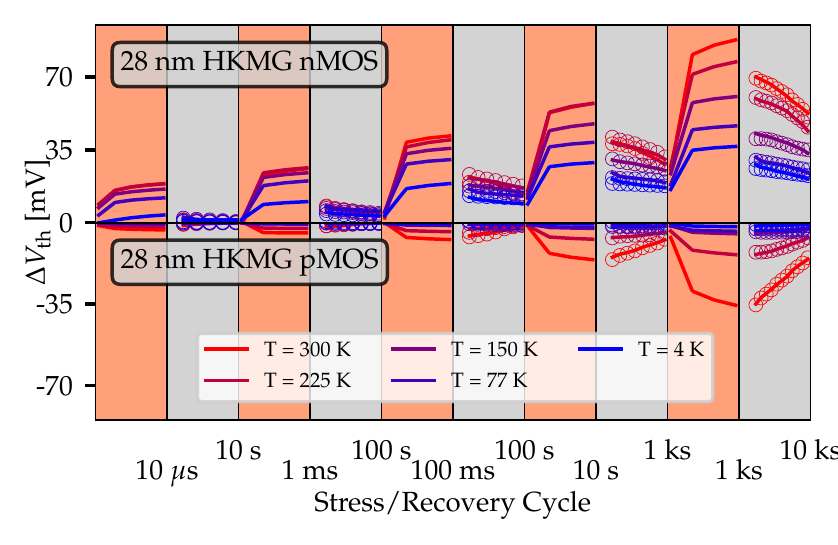}
	\caption{eMSM curves recorded between \SI{4}{\kelvin} and room temperature on a \SI{28}{\nm} HKMG technology show that the threshold voltage shift \dVth~of the pMOS device freezes out completely while the nMOS still shows a significant BTI shift. }
	\label{fig:cryo_dVth}
\end{figure}

\subsection{Cryogenic Modeling and ESiD}
Using these cryogenic models for charge transitions in combination with the ESiD framework presented in Sec.~\ref{sec:esid}, it is possible to extract trapbands corresponding to recorded eMSM-measurements. As shown for a \SI{28}{\nm}-process of a high-$\kappa$ metal-gate device, with a HfO$_2$ layer processed on a thin SiO$_2$ layer and dimensions $W\times L=\SI{10}{\micro\m}\times\SI{10}{\micro\m}$ in \reffig{fig:cryo_dVth}, the threshold voltage shift \dVth~after a stress bias gets smaller towards cryogenic temperatures. However, while for pMOS devices \dVth~freezes out completely below $T\approx \SI{150}{\kelvin}$ there is still a significant BTI shift on nMOS devices even at temperatures as low as $T=\SI{4}{\kelvin}$. 

\begin{figure}[t]
	\centering
	\includegraphics[width=\columnwidth]{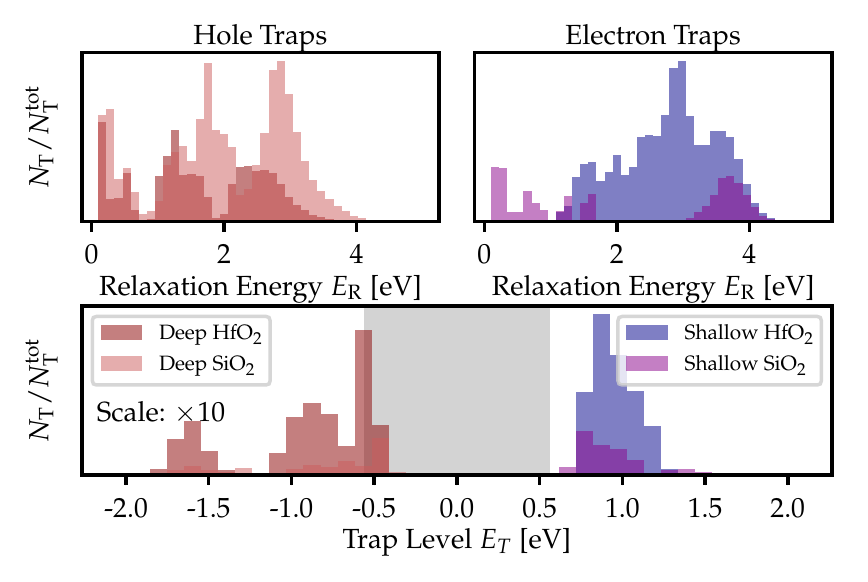}
	\caption{Distributions of trap levels \textbf{(top)} and relaxation energies \textbf{(bottom)} for electron and hole traps as extracted with the ESiD scheme. The Si bandgap is indicated by the gray area.}
	\label{fig:cryo_E_S_hist}
\end{figure}

Trapbands describing this asymmetry between nMOS and pMOS can be found using the ESiD-algorithm. The optimized trapbands describing the recorded \dVth~curves are shown in \reffig{fig:cryo_E_S_hist}. In line with earlier investigations based on CP experiments~\cite{kerber2003}, the trap distribution extracted with ESiD shows large amounts of electron traps in the HfO$_2$ layer. As discussed in \cite{Michl2021EfficientModelingOf2}, the nMOS-pMOS asymmetry at cryogenic temperatures can be explained by these electron traps in the HfO$_2$ layer and fast electronic interface traps located close to the Si conduction band edge.

\section{Outlook for Future Releases}

While these features already provide a substantial enhancement compared to the original version, future works will also include compact models for emerging 2D devices, refined electrostatic models at cryogenic temperatures, coupling of charge trapping and ferroelectricity for FeFET applications. In order to add more flexibility, a numerical Poisson-solver will be included as well to account for incomplete ionization and movement of possible mobile ions in the oxide. Moreover, it will be possible to efficiently study the impact of charge trapping on CV characteristics.

\section{Conclusions}
Device reliability becomes an ever increasing concern with the continued scaling into the atomic realm. Hence, considerable effort is put into developing predictive reliability models in order to assess the non-ideal behavior of nanoscale devices in simulations rather than costly and time-intensive experiments. Many prevalent reliability issues like bias temperature instability (BTI) or random telegraph noise (RTN) are dominated by charge trapping in oxides, demanding a solid understanding and a physical model for such charge transfer processes in devices. While the nonradiative multiphonon (NMP) theory provides a rigorous framework for the treatment of charge trapping, its implementation in commercial TCAD software is often slow and cumbersome to calibrate accurately.

Here, we have summarized our recent efforts to provide a lightweight realization of NMP theory coupled to a 1D compact device model, resulting in the release of \textit{Comphy~v3.0}, an open source Python package for reliability simulations. While the original \textit{Comphy~v1.0}~\cite{rzepa2018comphy} was intended as a proof-of-concept for simple but accurate reliability models, this new release~\cite{comphy3} provides a comprehensive framework to meet the current demands in industry and academic device research. Among the key features of \textit{Comphy v3.0} are \textit{(i)} an automated parameter extraction scheme which allows to extract defect parameters from simple extended measure-stress-measure (eMSM) experiments and to build predictive BTI models with ease; \textit{(ii)} a new unified approach to treat trap-assisted tunneling and BTI on an equal footing within NMP theory; and \textit{(iii)} a compact charge trapping model including nuclear tunneling at cryogenic temperatures  to study RTN and BTI in emerging fields like quantum computing. 

\section*{Acknowledgements}

The financial
support by the Austrian Federal Ministry for Digital and Economic
Affairs, the National Foundation for Research, Technology and
Development, the Christian Doppler Research Association, and the European Research Council (ERC) under grant agreement no.101055379 is
gratefully acknowledged.


\bibliographystyle{elsarticle-num}

\bibliography{bibliography}



\end{document}